\documentclass[varenna]{cimento}

\newcommand{\verysmallfig}{0.3\textwidth}
\newcommand{\smallfig}{0.5\textwidth}
\newcommand{\normfig}{0.7\textwidth}

\newcommand{\DD}{{\mathcal D}}
\newcommand{\Sel}{S_{\rm el}}
\newcommand{\Sdis}{S_{\rm dis}}
\newcommand{\Fvar}{S_{\rm var}}
\newcommand{\fref}[1]{Fig.~\ref{#1}}
\newcommand{\sref}[1]{Sec.~\ref{#1}}

\usepackage{graphicx}
\title{Electronic glasses}
\author{Thierry Giamarchi}
\institute{DPMC, Universite de Geneve, 24 Quai Ernest-Ansermet,
1211 Geneve 4, Switzerland}
\begin{document}

\maketitle

\section{Introduction}

For more than 50 years our understanding of interacting electronic
systems has been essentially based on the Fermi liquid theory. In
this way of viewing electrons, the electron is essentially
described by a (nearly free) plane wave excitation, with a well
defined momentum and energy (see e.g. \cite{nozieres_book}). Most
of the phenomena in solids can thus be understood in this wave
description, with the crucial addition of the Pauli principle
which embeds the antisymmetric nature of the total wavefunction of
the system. The Pauli principle blocks states too deep in energy
and ensures the existence of a Fermi surface. Such an approach has
been extremely successful to explain quite remarkable phenomena
such as superconductivity. It has been crucial also to interpret
the subtle effects of disorder in solids. Indeed although the
naive scattering of electrons on impurities can be understood in a
semi-classical approach, more subtle effects like Anderson
localization rests on the quantum nature of the particles and can
be interpreted as wave interferences. Using both scaling theories
and sophisticated field theoretical techniques, it is now known
that free electrons are localized by disorder in one and two
dimensions, whereas a mobility edge exists in three dimensions.
\cite{berezinskii_conductivity_log,abrikosov_rhyzkin,abrahams_loc,%
wegner_localisation,efetov_localisation,efetov_supersym_revue}

\begin{figure}
 \centerline{\includegraphics[width=\normfig]{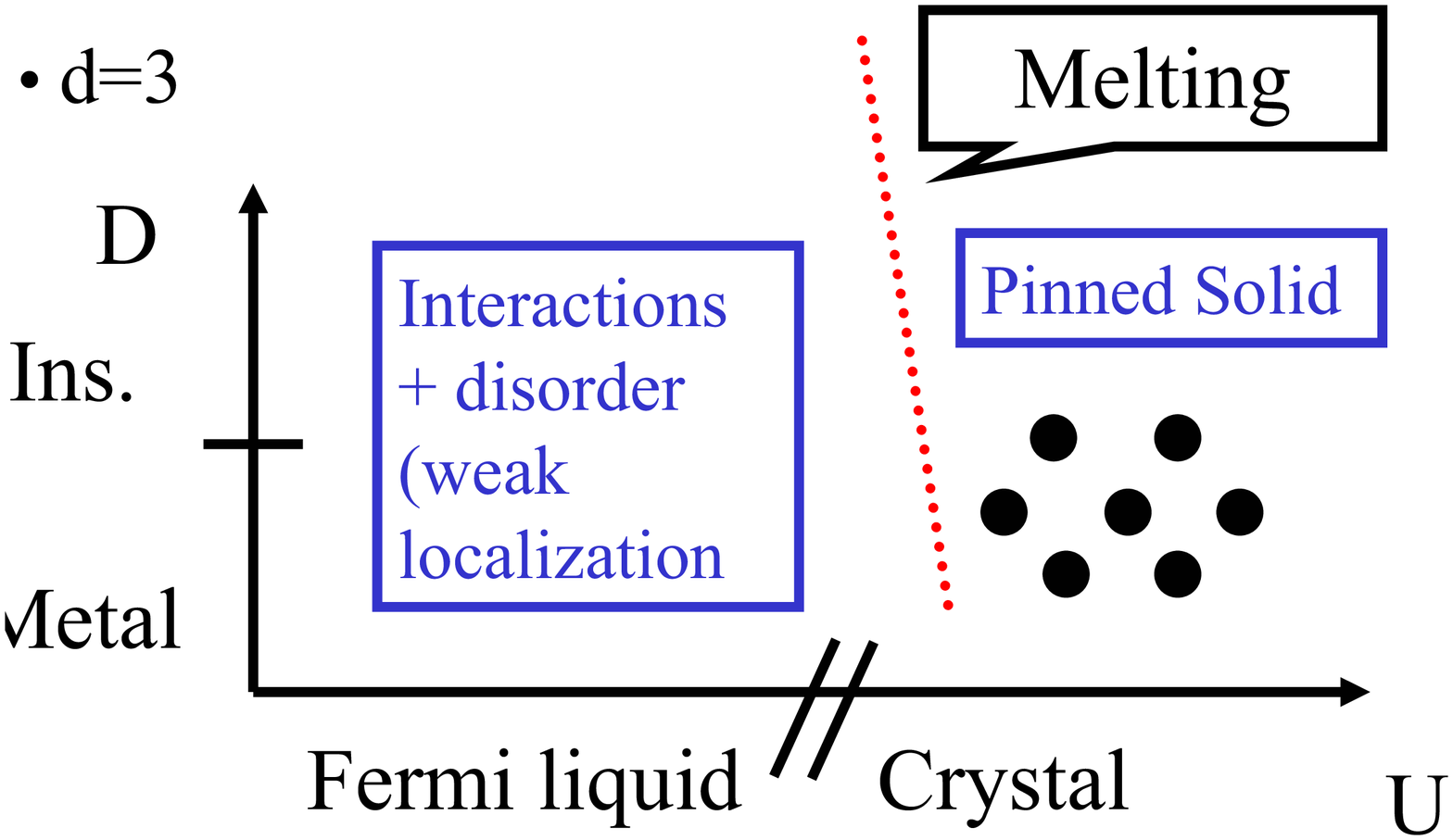}}
  \centerline{\includegraphics[width=\normfig]{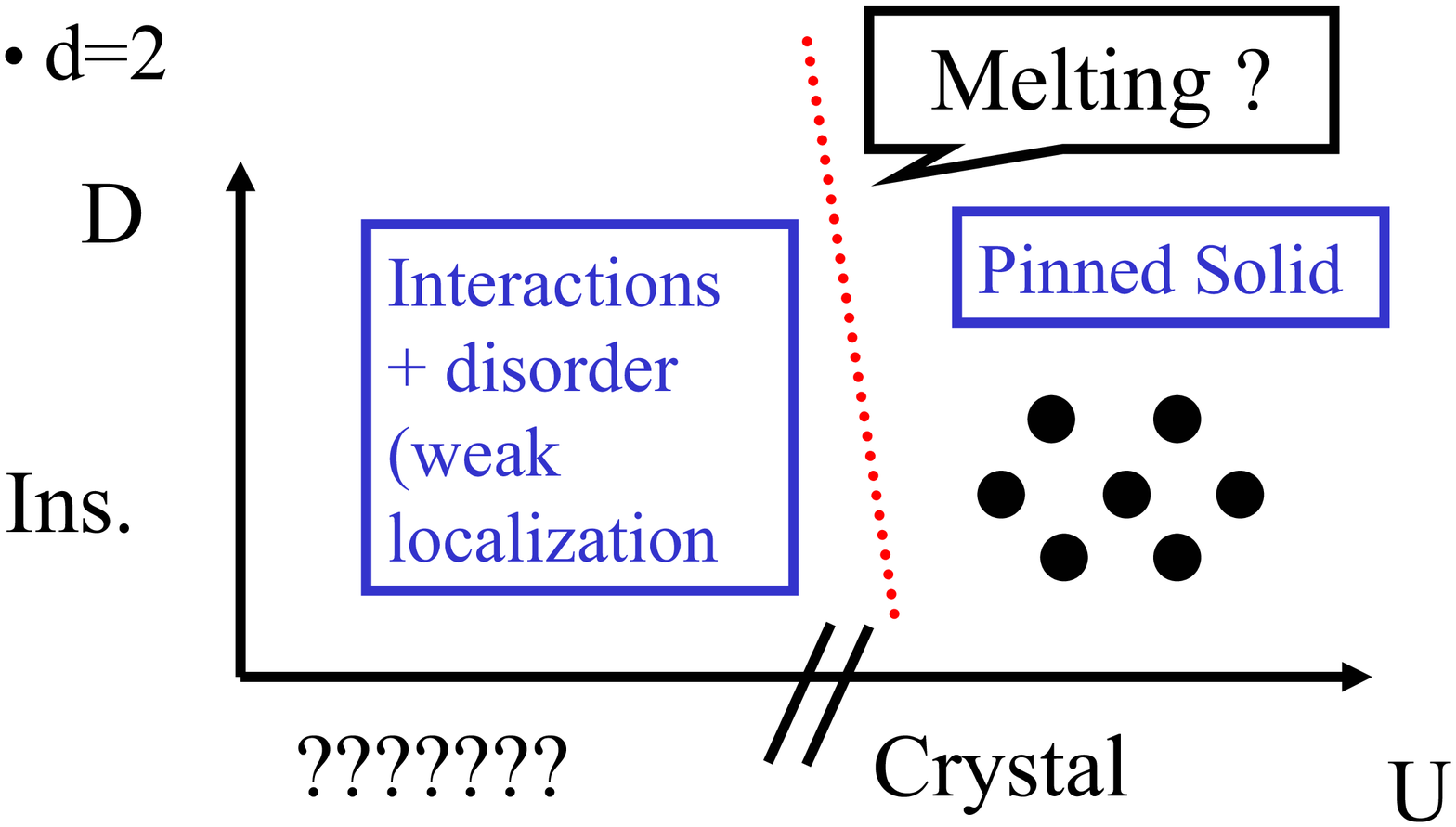}}
 \caption{(top) combined effects of disorder $D$ and interactions (denoted loosely $U$) in $d=3$.
 Whether an intermediate phase exists between the crystal phase and a
 standard Fermi liquid phase is an open question.
 (bot) same thing in two dimensions. In $d=2$ it is unclear whether a metallic phase exists, or whether
 a true crystalline phase can exist. Nevertheless, for strong interactions a description in terms of a crystal
 is a much better starting point than a plane wave description of the electrons. Although on these pictures the phase
 diagram is written in term of the interactions for simplicity,
 usually the density is varied or the kinetic energy is killed by application
 of a strong magnetic field in order to reach the crystal phase.}
 \label{fig:fermicrys}
\end{figure}

However when the interactions become strong this ``wave'' starting
point can be a poor description of the system. Indeed as was
described a long time ago by Wigner \cite{wigner_crystal}, if the
Coulomb interactions dominate the wave functions of each particle
becomes localized in space, instead of being a plane wave. The
electrons form a crystal (Wigner crystal (WC)). In such a crystal
phase our intuition based on the wave description of electrons
fails. The wavefunction of the electrons is localized so the
electrons become discernable particles by their position, and the
Pauli principle plays a much smaller role. Quantum fluctuations
are still there however since the vibration modes of such crystal
are still quantized. One is thus facing a novel phase of quantum
systems with quite fascinating properties to explore. For example,
for the pure system if the interactions are reduced the quantum
fluctuations will increase and one can expect the crystal to melt.
What is the nature of this melting and to which phase it melts is
of course a crucial, and still an open question. Whether this is
simply a Fermi liquid in which the particle have recovered their
``plane wave'' nature or whether some intermediate phase exists is
yet to be understood. Needless to say one can also expect quite
interesting effects of the disorder in such systems. In the
crystal phase the disorder effects cannot be viewed as scattering
of waves but rather as the pinning of this crystal on the random
potential due to impurities. It is thus interesting to know how
this pictures of a pinned crystal connects with the standard
description of interacting disordered systems, adding interactions
on the top of disordered free electrons
\cite{altshuler_aronov,finkelstein_localization_interactions,lee_mit_long}.
From a theoretical point of view this is quite a challenging
question. Most of the theoretical approaches used for free
electrons either fail or become much more complicated when
interactions are included which makes it more difficult to obtain
unambiguous answers. Perturbative calculations or renormalization
group calculations can be made for weak interactions.
Unfortunately they scale to strong coupling, which leaves the
question of the large scale/low energy physics still open. The
crystal description thus provides and alternative way to study
this quite complicated problem. Even in the case of the crystal
the competition of repulsion between particles, that wants the
crystal to be nicely ordered, and the impurities that want the
particles of the crystal at random position is a quite difficult
problem. As a results, physics has been mostly computed in the
limit of very strong disorder where particles are pinned
individually \cite{shklovskii_conductivity_coulomb}. A summary of
the situation is presented in \fref{fig:fermicrys}.

In addition to this pure theoretical interest, there is also a
direct experimental relevance. The progress of nanotechnology have
made it possible to produce such quantum crystals. There are
various situations where one can reach Wigner crystallization. In
order to enhance the interactions effects compared to the kinetic
energy it is necessary to reduce the electron density (kinetic
energy is killed faster than interactions at low densities). Thus
semiconductors where one can reach very low density of carriers
have been prime candidates to observe Wigner crystallization. Two
dimensional electron gas (2DEG) where the density can be
controlled by an external gate are good candidates. Because the
crystallization requires quite low densities, one can help it by
killing the kinetic energy of the electrons by application of a
very strong magnetic field. In that case the electrons describe
cyclotron orbits and their kinetic energy is quenched helping the
interactions. This allows to reach Wigner crystallization for
higher densities than in the absence of a field. Thus the first
signs of Wigner crystallization were observed in systems where the
fractional quantum hall effect was seen, but at higher magnetic
fields.

\begin{figure}
 \centerline{\includegraphics[width=\verysmallfig]{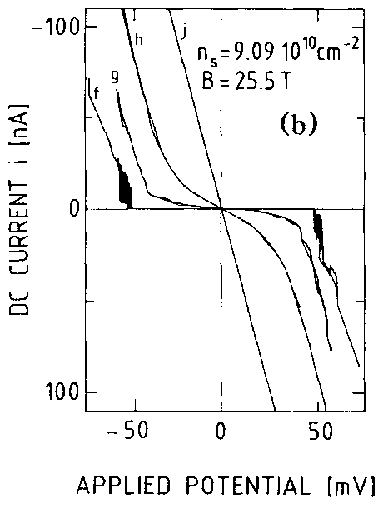}
             \includegraphics[width=\verysmallfig]{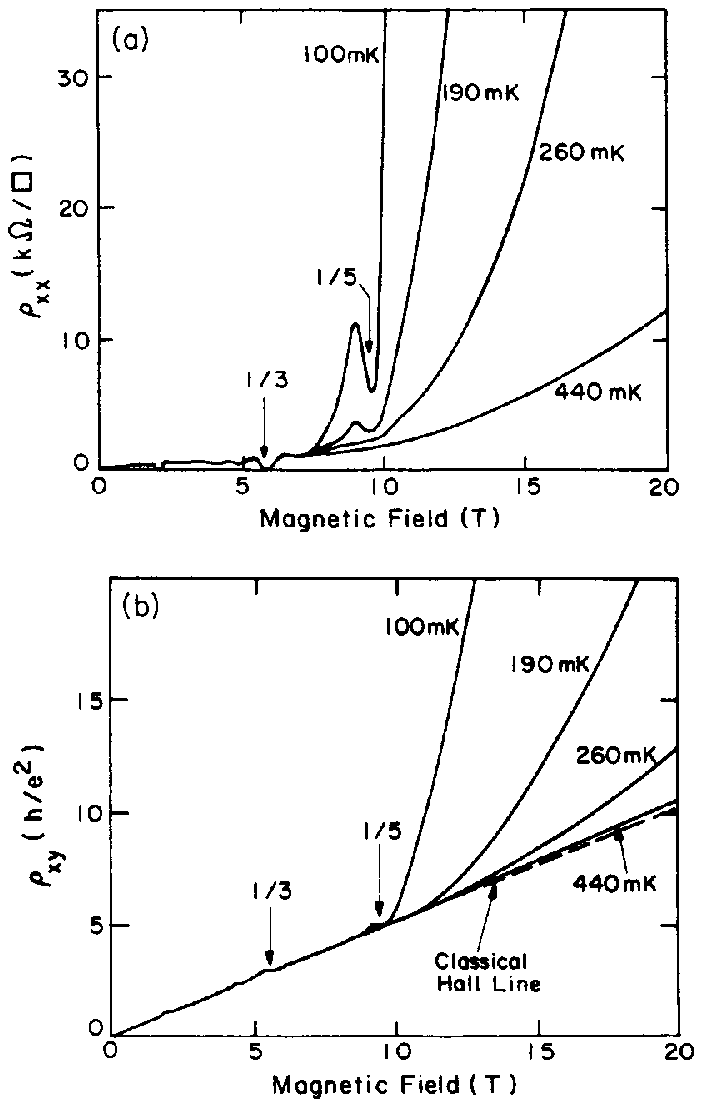}}
 \centerline{\includegraphics[width=\smallfig]{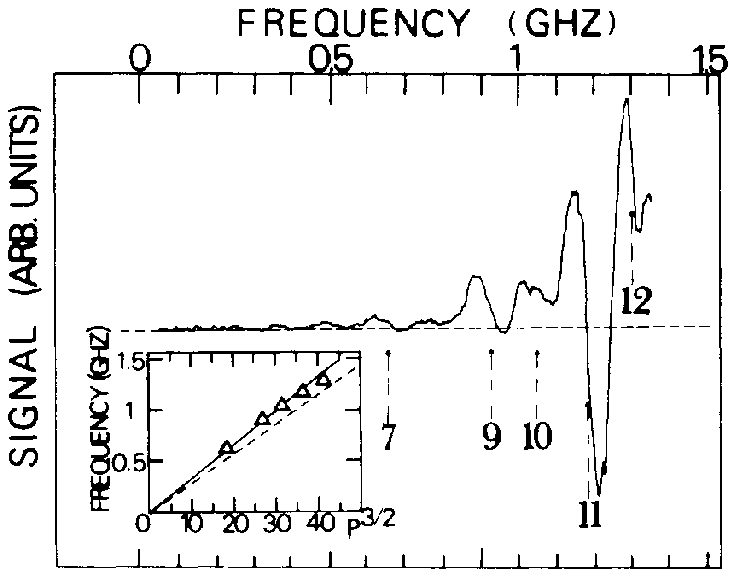}}
 \caption{(bottom) Soundwave absorption by a two dimensional
 electron gas (2DEG) under strong magnetic field. The frequencies
 at which the sound is absorbed correspond to the eigenmodes of the
 crystal (see \sref{sec:pure}), and are interpreted as evidence
 of a Wigner crystal in this system (from \cite{andrei_wigner_2d});
 (top Right) Transport properties of a 2DEG. At strong magnetic field
 an insulating phase appears, again suggestive of the formation of
 a pinned Wigner crystal (from \cite{willett_wigner_resistivity}).
 (top left) Current vs. voltage characteristics. One clearly sees
 a threshold field needed to have conduction. This is again
 reminiscent of what one expects of a pinned crystal (From
 \cite{williams_wigner_threshold}).}
 \label{fig:wignerfirst}
\end{figure}

Some experimental data in these systems is shown in
\fref{fig:wignerfirst}.
This data prompts for an immediate question: in all these systems
there is no direct evidence of the crystal structure. Some
attempts have been made to image such systems but one still cannot
resolve individual crystal sites
\cite{ilani_compressibility_2DEG}. This is at variance with
classical crystals such as the vortex lattice where good imaging
techniques exists (see e.g. \cite{giamarchi_vortex_review}), and
thus where it is possible to directly ``see'' the crystalline
order. Here one has really to infer the existence of the crystal
from circumstantial evidence, i.e. from transport measurements
(mostly) or some other indirect measurements (acoustic wave
absorption, compressibility etc.). In order to know whether the
transport experiments can be considered as a proof or not of the
existence of the Wigner crystal is it thus specially important to
have a reliable theory that allows to compute the transport
properties. Such a task is far from being trivial given the
complexity of the problem.

\begin{figure}
 \centerline{\includegraphics[width=\smallfig]{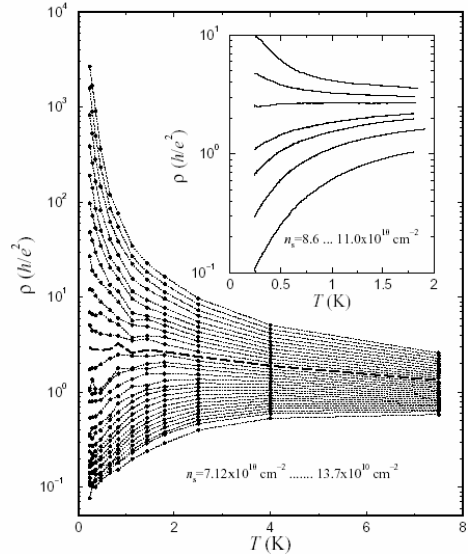}}
 \caption{Metal-insulator ``transition'' observed at zero magnetic field in 2DEG systems.
 Is the insulating phase (low density) a pinned Wigner crystal ? (From \cite{kravchenko_mit_scaling}).}
 \label{fig:kravchenko}
\end{figure}

Quite recently systems with even lower densities could be
achieved. In these systems a ``metal-insulator'' transition was
observed. There is considerable debate whether the metallic phase
really exists \cite{abrahams_review_mit_2d}, but in my sense a
more interesting question is whether or not the insulating phase
(which is the low density phase) is a pinned Wigner crystal.
Resistivity data is shown in \fref{fig:kravchenko}.
Finally and although I will not insist on these systems in these
notes, Luttinger liquids can be viewed also as an electronic
crystal. A detailed review of the properties of such systems and
their connection with disordered electronic crystals can be found
in \cite{giamarchi_quantum_pinning}.

There is thus considerable interest in understanding the
properties of these electronic crystals, which is the subject I
will present in these lectures. It is clear that I cannot expect
to cover such a subject in these few pages, so I will just give
the basic ideas and some references. From the theoretical side,
the concepts and methods used are quite different from the methods
used usually to tackle disordered electrons, and I will introduce
some of them. I will discuss the minimal model needed to describe
an electronic glass in \sref{sec:basics}. I will present some
basic concepts for disordered crystals in \sref{sec:con} and
discuss also some incorrect preconceived ideas on the physics of
such systems. In order to give a more precise solution I will
present two more sophisticated theoretical tools in
\sref{sec:methods}: the replica method and a functional
renormalization group. Some of the results for equilibrium
transport will be examined in \sref{sec:quantit}. Finally some
conclusions and perspectives can be found in
\sref{sec:conclusion}.

In complement and connection with the present notes there are
three other recent set of notes that present other aspects of
disordered elastic systems and where further references to other
reviews or original papers can be found. The notes
\cite{giamarchi_vortex_review} deal with classical crystals such
as vortex lattices. \cite{giamarchi_quantum_pinning} is a rather
complete review on quantum problems. It also presents other
techniques and discusses in details one dimensional systems. The
present notes are complementary with this review (there is some
overlap) in the sense that I will much more insist here on two
dimensional electronic systems. \cite{giamarchi_wigner_review}
deals with the specific case of Wigner crystal, and discusses in
much more detail the issues of compressibility and out of
equilibrium dynamics that I will only briefly mention in these
notes.

\section{Basic Description} \label{sec:basics}

\subsection{Elastic description}

\begin{figure}
\centerline{\includegraphics[width=\smallfig]{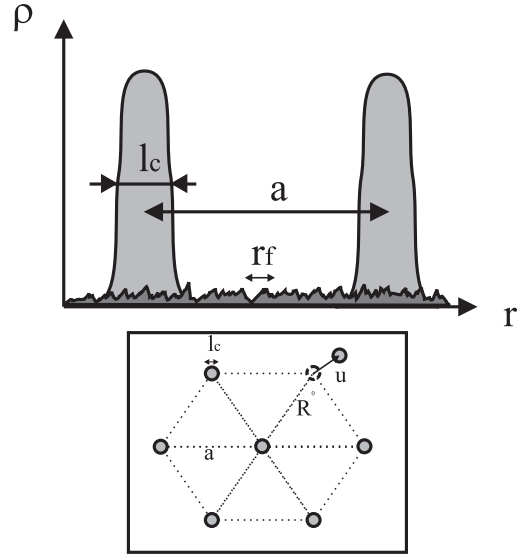}}
\caption{The three length characterizing the Wigner crystal. The
size $l_c$ of the ``particles'' in the crystal (at low temperature
it is essentially given by the extension of the wavefunction
around the equilibrium position, at large temperatures it is
controlled by the thermal fluctuations and is the Lindemann
length), $a$ the lattice spacing is controlled by the density of
particles, and the disorder is correlated over a length $r_f$. The
inset shows  the triangular structure of the Wigner crystal.
Particles are labeled by an equilibrium position $R^0_i$ and a
displacement $u_i$.(From \cite{chitra_wigner_long})}
\label{fig:basiclength}
\end{figure}

\begin{figure}
 \centerline{\includegraphics[width=\normfig]{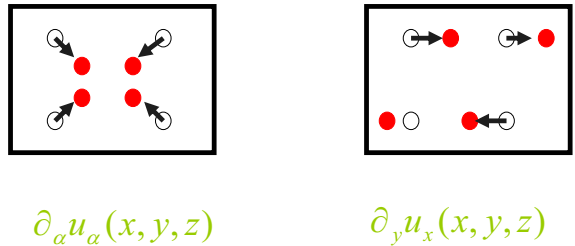}}
 \caption{(left) Bulk compression mode. This mode changes the
 smeared density and is thus coupled to the long range nature of
 the interaction. (right) shear mode. This mode leaves the smeared
 density invariant and feels only the short range part of the
 interaction.}
 \label{fig:bulkshear}
\end{figure}

Starting from the full electronic Hamiltonian (fermions with
interactions and disorder) is a near impossible task. In the
crystal phase some simplifications can be made since the particles
are now discernable by their position. This allows for a minimal
phenomenological model to describe such a crystal
\cite{giamarchi_columnar_variat,chitra_wigner_hall,chitra_wigner_long,giamarchi_wigner_review}:
one assumes that the particles are characterized by an equilibrium
position $R^0_i$ and a displacement $u_i$ relative to this
equilibrium position. In order to define uniquely the displacement
one should not have topological defects such as dislocations in
the crystal. I will come back to this point in \sref{sec:defects}.
From the original quantum problem one has to define the
``particles'' of the crystal. If the wavefunction is localized
enough then one can indeed ignore the exchange between the various
sites and thus define ``particles'' that have a size given by the
extension of the wavefunction as indicated in
\fref{fig:basiclength}.
Of course the density fixes the lattice spacing $a$. These two
lengthscales (size of particle, lattice spacing) are independent
and should be kept. One then has to deal with the vibration modes
of this crystal. As usual one considers the interaction between
particles
\begin{equation}
 H = \frac12 \sum_{i\ne j} V(R_i - R_j) = \frac12\int dr dr'
 V(r-r')\rho(r)\rho(r')
\end{equation}
and expands to second order in the displacements. If the potential
$V$ is short range this gives the standard elasticity. Here one
has to be more careful since $V$ is the (long range) Coulomb
interaction between the electrons. One has thus to distinguish the
two modes presented in \fref{fig:bulkshear}.
The compression mode changes the smeared density at one point of
the crystal, creating an excess or a default charge. Such charges
interact through the long range part of the Coulomb interaction.
Since the density is essentially $\nabla u$ (I will be more
precise later), this mode gives a contribution $V(q) q^2 u^*_q
u_q$ in the energy, where $V(q)$ is the Fourier transform of the
Coulomb interaction. Since $V(r) \sim 1/r$, $V(q) \sim 1/q$ in two
dimensions. The compression mode thus has an elasticity going as
$|q| u^*_q u_q$. On the other hand the shear mode does not change
the density (averaged over a couple of unit cells), and thus the
long range part of the interaction does not couple to this mode.
The coupling comes from the short range part (or if one prefers
$V(q \sim 1/a)$). The shear mode has thus still an elasticity
going as $q^2$. Being more precise one can define the longitudinal
and transverse displacements for a mode $q$ by
\begin{equation}
 \vec{u}(q) = \frac{\vec{q}}{q}
  u_L(q) + (\frac{\vec{q}}{q} \wedge \vec{z}) u_T(q)
\end{equation}
This leads to the action (see \cite{chitra_wigner_long} for more
details):
\begin{eqnarray} \label{eq:ham}
 S &=& \frac12 \int_{\bf  q} \frac1\beta \sum_{\omega_n}
 u_{L}(q,\omega_n)(\rho_m\omega_n^2 +c_L(q))u_{L}(-q,-\omega_n) \\
 & & + u_{T}(q,\omega_n)(\rho_m\omega_n^2 +c_T(q))u_{T}(-q,-\omega_n) \nonumber \\
 & & + \rho_m \omega_c \omega_n [u_{L}(q,\omega_n)u_{T}(-q,-\omega_n)-
 u_{T}(q,\omega_n)u_{L}(-q,-\omega_n)] \nonumber
\nonumber
\end{eqnarray}
$\int_{\bf q}$ denote the integration over the Brillouin zone
$\int_{BZ} \frac{d^2q}{(2\pi)^2}$ and the $\omega_n$ are the
standard Matsubara frequencies. The $\omega_n^2$ terms are coming
from the $P^2/2m$ term in the Hamiltonian, and are the ones
embedding the quantum nature of the crystal. For a quantum crystal
it is impossible to have the particles at rest since both its
position and position would be determined simultaneously. Thus
even at $T=0$ quantum fluctuations exists. The third term in
(\ref{eq:ham}) comes from the Lorentz force, that exists if the
particles are submitted to a magnetic field. The action
corresponding to this force is simply the work of the force $F_L
\cdot u$ where $F_L = q \partial_\tau u \wedge B$. $\rho_m \simeq
{m \over {\pi a^2}}$ is the mass density and $\omega_c = e B/m$
the cyclotron frequency.

The precise coefficients $C_{L,T}(q)$ are the elastic coefficients
for the longitudinal and transverse modes respectively. These
coefficients can be obtained from an expansion of the coulomb
correlation energy of the WC in terms of the displacements
\cite{bonsall_elastic_wigner,maki_elastic_wigner}. As I discussed
$C_L(q) \sim |q|$ and $C_T(q) \sim q^2$. Keeping the lowest terms
of the expansion one has
\begin{eqnarray}
 C_L(q) &=& d |q| - b q^2 \nonumber \\
 C_T(q) &=& c q^2
\end{eqnarray}
where $c$,$d$,$b$ are positive constants.

\subsection{Consequences for pure system} \label{sec:pure}

For a pure system the consequences of the quadratic Hamiltonian
(\ref{eq:ham}) are easy to carry out. The eigenmodes of the system
are obtained by doing the analytic continuation $i\omega_n \to
\omega + i\delta$. In the absence of magnetic field the
longitudinal and transverse modes are decoupled. The longitudinal
one has a dispersion
\begin{equation}
 \rho_m \omega^2_L = C_L(q) \sim d|q|
\end{equation}
This is the standard plasmon dispersion in $d=2$. Because of the
Coulomb repulsion it is extremely hard to excite this mode. The
system is very rigid (much more than a normal elastic system)
since it does not like to have its local density changed. This is
not the case of the transverse mode which keeps a phonon-like
dispersion
\begin{equation}
 \rho_m \omega^2_T = C_T(q) \sim c q^2
\end{equation}

In the presence of a large magnetic field the two modes are mixed.
It is easy to obtain the new eigenmodes by diagonalizing the two
by two matrix in (\ref{eq:ham}). This gives the modes of
Table~\ref{tab:modes}.
\begin{table}
\begin{tabular}{c|c|c}
 \hline
 Mode & zero field & High magnetic field \\
 \hline
 $\omega_-(q)$ & $\propto q^2$ (trans.) & $\propto \frac{q^{3/2}}{\omega_c}$ \\
 $\omega_+(q)$ & $\propto q  $ (long.)  & $\sim  \omega_c$ \\
 \hline
\end{tabular}
\caption{$q$ dependence of the eigenmodes in a Wigner crystal in
the absence of magnetic field or for a very strong
field.}\label{tab:modes}
\end{table}
The ``plasmon'' mode is now pushed to the cyclotron frequency
$\omega_c$, while there is a low energy mode with a $q^{3/2}$
dispersion. This low energy mode is the mode that was probed in
the sound absorption experiment shown in \fref{fig:wignerfirst}.

For the crystal, the current is simply given by $J = e \rho_0
\partial_t u$, making thus the conductivity very simple to compute
since it is essentially the correlator of the displacements (up to
a factor $\omega^2$). The conductivity is thus given by
\begin{equation}
 \sigma_{\alpha\beta}(\omega) = -(e \rho_0)^2 i(\omega+i\delta) \langle
 u_\alpha(q=0,\omega_n)^*u_\beta(q=0,\omega_n) \rangle_{i\omega_n\to\omega+i\delta}
\end{equation}
In the absence of magnetic field the conductivity is simply
\begin{equation}
 \sigma_{xx}(\omega) = \frac{e^2\rho_0}{m}\frac{i}{\omega+i\delta}
\end{equation}
The real part is thus just a $\delta$-function at zero frequency.
This traduces the fact that the crystal can slide freely in
response to an external force (the electric field). The pure
quantum crystal is thus a perfect conductor. In the presence of a
finite magnetic field, the conductivity becomes
\begin{equation}
 \sigma_{xx}(\omega) = \sigma_{yy}(\omega) =
 \frac{e^2\rho_0}{m}\frac{-i(\omega+i\delta)}{-(\omega+i\delta)^2 + \omega_c^2}
\end{equation}
Thus the $\delta$-function peak is pushed at the finite frequency
$\omega_c$. The d.c. conductivity is zero. This is due to the fact
that in presence of a magnetic field the electrons describe
cyclotron orbits, and do not drift in response to a static
electric field. Paradoxically the system is then an insulator. On
the other hand the Hall conductivity is given by
\begin{equation}
 \sigma_{xy}(\omega) = \frac{e^2\rho_0}{m}\frac{\omega_c}{-(\omega+i\delta)^2 + \omega_c^2}
\end{equation}
This gives for the Hall resistivity
\begin{equation}
 \rho_{xy}(\omega) = \frac{\sigma_{xy}}{\sigma_{xx}^2+\sigma_{xy}^2}
 = \frac{m \omega_c}{e^2 \rho_0} = \frac{B}{e \rho_0}
\end{equation}
which quite remarkably is the \emph{classical} Hall resistivity.

Of course these results are for the pure system only and the
crucial question, in order to make contact with the experiments,
is to determine how the disorder changes the above results.

\subsection{Disorder}

The coupling of the crystal to the disorder writes
\begin{equation} \label{eq:sdis}
 \Sdis = \int d\tau \int dr V(r) \rho(r,\tau)
\end{equation}
The potential $V$ represents the impurities. In these systems
there are usually two main sources of disorder. One is provided by
donors/acceptors far from the 2DEG. They interact with the 2DEG
via the long range Coulomb interaction (see e.g.
\cite{rusin_shklovskii_wigner}). Since they are far from the 2DEG
this produces a smooth potential. Another source of disorder is
provided by interface roughness \cite{williams_private_disorder}.
This type of disorder has a wavelength which is comparable or
shorter than the lattice spacing of the crystal, and is most
likely the main source of pinning.

\begin{figure}
\centerline{\includegraphics[width=\normfig]{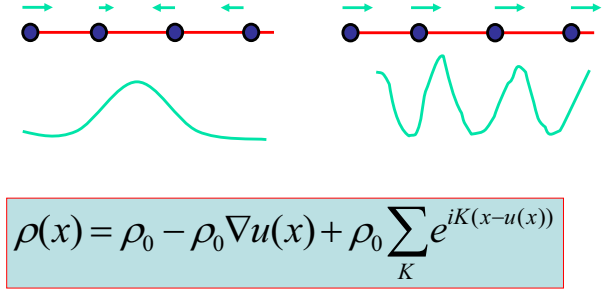}}
\caption{\label{fig:decdens} Various harmonics of the density. If
one is only interesting in variations of the density at
lengthscales large compared to the lattice spacing one has the
standard ``elastic'' expression of the density in terms of the
displacements. In a crystal however on has to consider variations
of density at lengthscales smaller than the lattice spacing and
higher harmonics are needed
\cite{giamarchi_vortex_short,giamarchi_vortex_long}.}
\end{figure}

In the limit where the impurities are weak and the pinning is
collective, the disorder can be modelled by a gaussian potential,
(see \cite{giamarchi_columnar_variat,giamarchi_vortex_review} for
a more in depth discussion of this limit)
\begin{equation}
\overline{V(r) V(r')} = \Delta_{r_f}({ r}-{ r}')
\end{equation}
$\Delta_{r_f}$ is a delta-like function of range $r_f$ which is
the characteristic correlation length of the disorder potential
(see Figure~\ref{fig:basiclength}). The density reads
\begin{equation} \label{eq:deldens}
\rho(r) = \sum_i \overline{\delta}(r - R^0_i - u_i)
\end{equation}
where $\overline{\delta}$ is a $\delta$-like function of range
$l_c$ (see Figure~\ref{fig:basiclength}) and $u_i \equiv
u(R^0_i)$. Since the disorder can vary at a lengthscale $r_f$ {\it
a priori} shorter or comparable to the lattice spacing $a$, the
continuum limit $u_i \to u(r)$, valid in the elastic limit $|u_i -
u_{i+1}| \ll a$ should be taken with care in the disorder term.
This can be done using the decomposition of the density in terms
of its Fourier components
\cite{giamarchi_vortex_short,giamarchi_vortex_long}
\begin{equation} \label{eq:fourdens}
 \rho(r) \simeq \rho_0 - \rho_0\nabla\cdot { u} + \rho_0 \sum_{{
 K} \neq 0} e^{i { K}\cdot({ r} - { u}(r))}
\end{equation}
where $\rho_0$ is the average density and $K$ are the reciprocal
lattice vectors  of the perfect crystal. This is a decomposition
of the density in Fourier harmonics determined by the periodicity
of the underlying perfect crystal as shown on \fref{fig:decdens}.
Let me examine the various terms. The first term is just the
average density. The second one is the textbook expression of the
density in terms of the displacements. It corresponds to a density
averaged at a scale of a few lattice spacing (smeared density).
This part of the density thus traduces local changes of the
density due to compression modes similar to the ones in
\fref{fig:bulkshear}. The other terms describes change of the
density at a lengthscale comparable or \emph{smaller} than the
lattice spacing. The sum over the reciprocal lattice vectors $K$
obviously reproduces the $\delta$ function peaks of the density
(\ref{eq:deldens}). The finite range of $\overline{\delta}$ is
recovered \cite{giamarchi_vortex_long} by restricting the sum over
$K$ to momentum of order $K_{\rm max} \sim \pi/l_c$. This
expression of the density is quite useful since it allows to
express the density at \emph{all} scales in term of the slowly
varying function $u(r)$. The price to pay is the more complicated
form of the operator that embeds in its very form the information
about the periodicity of the crystal and the size of the
particles. This is a quite useful trick, initially introduced for
Luttinger liquids \cite{haldane_bosons}. One keeps only the low
energy-long wavelength part of the displacement field, since this
field has a well defined continuum limit, and thus a quite simple
continuum Hamiltonian (elastic Hamiltonian in this case). But it
is still possible using this field that varies at a lengthscale
large compared to the lattice spacing (this is the essence of the
elastic limit) to describe phenomena that vary at a much shorter
lengthscale, provided that one uses the proper form for the
operator in terms of this continuum field. This expression of the
density allows to write quite simply the coupling to disorder.
Using the representation of density (\ref{eq:fourdens}) it is easy
to show \cite{giamarchi_columnar_variat,giamarchi_quantum_pinning}
that a Luttinger liquid is indeed very similar to the quantum
crystals that are described here.

Using the expression of the density (\ref{eq:fourdens}) one sees
that different Fourier component of the disorder play different
roles. The Fourier components close to $q\sim 0$ couple to $\nabla
u$. These terms play little role (see
\cite{giamarchi_quantum_pinning} for more details). In our case
because the compression modes are quite stiff thanks to the
Coulomb repulsion this is even more so. The main effect of the
disorder comes from the other Fourier modes, i.e. the ones that
are close to $q\sim K$ where $K$ is one of the vectors of
reciprocal lattice of the crystal. The effect of the disorder can
thus be written as
\begin{equation}
 \Sdis \simeq \sum_{K\ne 0} \int dr V(r)\rho_0 e^{iK(r-u(r))}
\end{equation}
The coupling to the displacements is highly nonlinear which of
course makes this problem very difficult to solve. The action
(\ref{eq:sdis}) and (\ref{eq:ham}) is the minimal model needed to
describe a disordered quantum crystal.

\begin{figure}
 \centerline{\includegraphics[width=\smallfig]{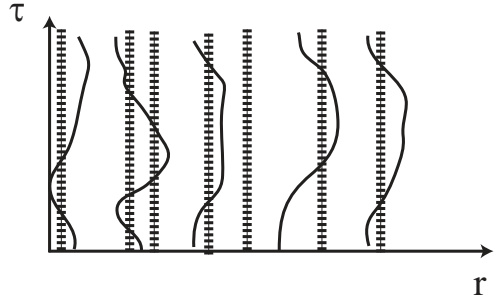}}
 \caption{The quantum crytal can also be
 viewed as a periodic array of lines that get pinned by columnar disorder. i.e. by disorder
 that is independent of one of the coordinates $\tau$ which is the imaginary time for
 the quantum problem. The lines are the space-time trajectories of the particles. Although the
 action can be computed with the imaginary time $\tau$, to obtain
 the \emph{physical} response for the quantum problem an analytic continuation is needed.
 This analytical continuation has an impact on many physical quantities, so the analogy between
 the classical system and the quantum one should be exerted with care.}
 \label{fig:disanal}
\end{figure}

For a quantum problem, the disorder is time independent. Thus if
one looks at the space-time trajectories of the particles, as
shown in \fref{fig:disanal}, one can view the system as a
classical system of lines in $d+1$ dimensions
\cite{giamarchi_columnar_variat,giamarchi_quantum_pinning}. These
lines are pinned by columnar defects, i.e. defects independent of
one of the coordinates ($\tau$ in this case).
The problem of lines pinned by columnar defects has been studied
extensively in the context of vortex lattices
\cite{nelson_columnar_long,blatter_vortex_review}, where in that
case the defects are produced by heavy ions irradiation. Thus some
features of the two systems are indeed similar and one can gain
considerable insight by thinking in term of the classical system
of lines. There is however an important difference. In the
classical system $\tau$ is simply one spatial direction, one is
thus looking at the static of such systems. The dynamics of
classical systems is a different problem, where an additional time
$t$ needs to be introduced (see e.g.
\cite{giamarchi_vortex_review,nattermann_vortex_review}). For the
quantum problem, $\tau$ is the \emph{imaginary time} of the
problem. Thus $\tau$ dependence of the quantities is describing
the \emph{equilibrium dynamics} of the quantum problem. In order
to get the proper quantum dynamics however one should make the
analytical continuation $\tau \to i t$ (or more precisely
$i\omega_n\to \omega+i\delta$). This analytical continuation,
although seemingly innocent has severe consequences and changes
many behaviors. Though although some analogy between the
\emph{static} (but $z=\tau$ dependent quantities such as e.g. the
tilt of the lines) behavior of classical crystals with columnar
defects and the \emph{dynamics} behavior of quantum crystals can
be made, one should exert great care because of the analytic
continuation needed for the quantum problems, which ultimately
leads to some differences in physics.

\section{Basic description and preconceived ideas} \label{sec:con}

\subsection{Relevance of disorder and basic lengthscales}

In order to define the basic properties of such systems let me
consider first what happens in the ground state of the problem.
Since the disorder is time independent, one can consider all
displacements as time independent to get the ground state (it
corresponds to the mode $\omega_n = 0$). The problem is thus
equivalent to a classical system in $d$ dimensions. The time
integration is just providing a factor $\beta$ in front of the
action and the problem is equivalent to determine the partition
function of a systems
\begin{equation}
 Z = \int \DD u[r] e^{- \beta H_{cl}}
\end{equation}
where
\begin{equation} \label{eq:hamclass}
 H_{\rm cl} = \frac{c}{2} \int d^dr (\nabla u)^2  + \rho_0 \int V(r)\rho(r)
\end{equation}
I have forgotten for simplicity in the above formula the
difference between longitudinal and transverse modes and simply
rewritten everything with a standard elasticity (this corresponds
in fact to the transverse mode with an elastic constant $C_T(q)
\sim q^2$). It is thus enough to examine a classical crystal, in
presence of point like defects (for more details see
\cite{giamarchi_vortex_review,nattermann_vortex_review}).

In order to understand the basic physics of such problem a simple
(but ground breaking !) scaling argument was put forward by Larkin
\cite{larkin_70}. To know whether the disorder is relevant and
destroys the perfect crystalline order, let me assume that there
exists a characteristic lengthscale $R_a$ for which the relative
displacements are of the order of the lattice spacing $u(R_a) -
u(0) \sim a$. If the displacements vary of order $a$ over the
lengthscale $R_a$ the cost in elastic energy from
(\ref{eq:hamclass}) is
\begin{equation} \label{eq:loss}
 \frac{c}2 R_a^{d-2} a^2
\end{equation}
by simple scaling analysis. Thus in the absence of disorder
minimizing the energy would lead to $R_a =\infty$ and thus to a
perfect crystal. In presence of the disorder the fact that
displacements can adjust to take advantage of the pinning center
on a volume of size $R_a^d$ allow to gain some energy. Since $V$
is random the energy gained by adapting to the random potential is
the square root of the potential over the volume $R_a^d$, thus one
gains an energy from (\ref{eq:hamclass})
\begin{equation} \label{eq:gain}
- V R_a^{d/2} \rho_0
\end{equation}
Thus minimizing (\ref{eq:loss}) plus (\ref{eq:gain}) shows that
below four dimensions the disorder is always relevant and leads to
a finite lengthscale
\begin{equation} \label{eq:larkinlength}
 R_a \sim a \left(\frac{c^2 a^d}{V^2 \rho_0^2}\right)^{1/(4-d)}
\end{equation}
at which the displacements are of order $a$. The conclusion is
thus that even an arbitrarily weak disorder destroys the perfect
positional order below four spatial dimensions, and thus no
disordered crystal can exist for $d \leq 4$. This is an
astonishing result, which has been rediscovered in other context
(for charge density waves $R_a$ is known as Fukuyama-Lee
\cite{fukuyama_pinning} length and for random field Ising model
this is the Imry-Ma length \cite{imry_ma}). Of course it
immediately prompt the question of what is the resulting phase of
elastic system plus disorder?

Since solving the full problem is tough another important step was
made by Larkin \cite{larkin_70,larkin_ovchinnikov_pinning}. For
small displacements he realized that (\ref{eq:hamclass}) could be
expanded in powers of $u$ leading to the simpler disorder term
\begin{equation}
 H_{\rm larkin} = \int d^dr f(r) u(r)
\end{equation}
where $f(r)$ is some random force acting on the vortices. Because
the coupling to disorder is now linear in the displacements the
Larkin Hamiltonian is exactly solvable. Taking a local random
force $\overline{f(r)f(r')} = \Delta \delta(r-r')$ gives for the
relative displacements correlation function and structure factor
\begin{eqnarray}
 B(r) = \overline{\langle[u(r)-u(0)]\rangle} &=& B_{\rm thermal}(r) +
 \frac{\Delta}{c^2} r^{4-d}
\end{eqnarray}
where $\langle\rangle$ denotes the thermodynamics average and
$\overline{\cdots}$ denotes the disorder average. $B_{\rm
thermal}$ is the same correlation function in the absence of
disorder due to thermal fluctuations (which remain bounded in
$d>2$ and are thus negligeable at large distance compared to the
disorder term). Thus the solution of the Larkin model confirms the
scaling analysis: (i) displacements do grow unboundedly and thus
perfect positional of the crystal is lost; (ii) the lengthscale at
which the displacements are of the order of the lattice is the
similar to the one given by the scaling analysis. In addition the
Larkin model tells us how fast the positional order is destroyed:
the displacements grow as power law thus one can expect the
positional order to be  destroyed exponentially fast. However
these two conclusions should be taken with a serious grain of
salt. Indeed the Larkin model is an expansion in powers of $u$,
and thus cannot be valid at large distance (since the
displacements grow unboundedly the expansion has to break down at
some lengthscale). What is this characteristic lengthscale? A
naive expectation is that the Larkin model cease to be valid when
the displacements are of order $a$ i.e. at $r=R_a$. In fact this
is too naive as was understood by Larkin and Ovchinikov. To
understand why, in a transparent way, let me use the expression
for the density (\ref{eq:fourdens}) This immediately allows to
reproduce the Larkin model by expanding the disorder term
(\ref{eq:hamclass})
\begin{equation}
 \rho_0 \int d^d r \sum_K e^{i K(r - u(r))} V(r)
\end{equation}
in powers of $u$. Clearly the expansion is valid as long as
$K_{max} u \ll 1$ This will thus be valid up to a lengthscale
$R_c$ such that $u(R_c)$ is of the order of the {\it size of the
particles} $l_c$. Note that this lengthscale is different (and
quite generally smaller) than the lengthscale $R_a$ at which the
displacements are of the order of the lattice spacing $a$. The
Larkin model cease to be valid way before the displacements become
of the order of $a$ and thus {\it cannot be used} to deduce the
behavior of the positional order at large length scale. In
addition it is easy to check that because the coupling to disorder
is linear in the Larkin model, this model does not exhibit any
pinning, since the model is invariant by changing $u(r) \to u(r) +
C$ where $C$ is an arbitrary constant (remember that the total of
the random forces has to be zero $\int dr f(r) = 0$). Thus sliding
by an arbitrary amount does not cost any energy and any addition
of an external force leads to a sliding of the vortex lattice.
Thus the lengthscale at which this model breaks down is precisely
the lengthscale at which pinning appears
\cite{larkin_ovchinnikov_pinning}. The lengthscale $R_c$ is thus
the lengthscale above which various chunks of the system are
collectively pinned by the disorder. A simple scaling analysis on
the energy gained when putting an external force
\begin{equation}
 H = \int d^dr F_{\rm ext} u(r)
\end{equation}
allows to determine the critical force needed to unpin the
lattice. Assuming that the critical force needed to unpin the
lattice is when the energy gained by moving due to the external
force is equal to the balance of elastic energy and disorder
$\frac{c}2 l_c^2 R_c^{d-2}$, one obtains for the critical force
needed to move the lattice
\begin{equation} \label{eq:lor}
 F_c \propto \frac{c l_c}{R_c^2}
\end{equation}
This is the famous Larkin-Ovchinnikov relation which allows to
relate a dynamical quantity (the critical force at $T=0$ needed to
unpin the lattice) to purely static lengthscales, here the
Larkin-Ovchinikov length at which the displacements are of the
order of the size of the particle. Let me insist again that this
lengthscale controlling pinning is quite different from the one
$R_a$ at which displacements are of the order of the lattice
spacing at that controls the properties of the positional order.

These two lengthscales $R_c$ and $R_a$ are thus the crucial ones
in controlling the properties of a pinned classical crystal, and
thus will of course be crucial for our quantum crystals. For the
quantum crystals this leaves us with two main questions: (i) what
is the conductivity of the system, and which length controls it;
(ii) what happens beyond the length $R_a$.

\subsection{Conventional wisdom and dynamics} \label{sec:wisdom}

Up to know we have been on relatively firm ground, but answering
the above questions, even for a classical crystal is a tough
cookie. Given the difficulty in obtaining the answers, instead of
a solution people have been happy for a long time with a kind of
consensus on how the system ``should'' behave, based on physical
intuition and some incomplete, or in some cases incorrect proofs.
I will give in this section a summary of these preconceived ideas
on disordered crystals, since they were at the root of most of the
physical solutions given for the dynamics. Fortunately recent
progress have made it possible to truly analyze these systems. As
usual the true behavior turned out to be quite different from what
was naively believed. I will examine these solutions in the next
section. But for the moment let me go along and describe the
(incorrect) picture of a disordered crystal, that was used for a
long time.

\begin{figure}
 \centerline{\includegraphics[width=\normfig]{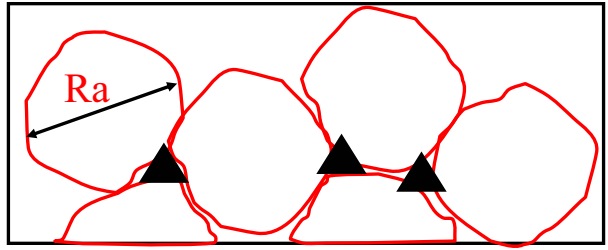}}
 \caption{The (incorrect) physical image
 that was the commonly accepted view of what a disordered elastic
 system would look like. The crystal would be broken into
 crystallites of size $R_a$ by the disorder. Dislocations would be
 generated by the disorder at the same lengthscale.}
 \label{fig:crystallite}
\end{figure}

The first point is to naturally assume that because there exists a
lengthscale $R_a$ for which displacements are of the order of the
lattice spacing, positional order is destroyed exponentially fast
beyond this lengthscale. This behavior would be in agreement with
the too naive extrapolation of the Larkin model to arbitrary
lengthscales and is in agreement with exact solutions of interface
problems in random environments (so called random manifold
problems) and solutions in one spatial dimension. It was thus
quite naturally assumed that an algebraic growth of displacements
was the correct physical solution of the problem, and thus that
the positional order would be destroyed exponentially beyond the
length $R_a$. This led to an image of the disordered vortex
lattice that consisted of a crystal ``broken'' into crystallites
of size $R_a$ due to disorder. To reinforce this image (incorrect)
``proofs'' were given \cite{fisher_vortexglass_long} to show that
due to disorder dislocations would be generated at the lengthscale
$R_a$ (even at $T=0$) further breaking the crystal apart and
leaving no hope of keeping positional order beyond $R_a$. A
summary of this (incorrect) physical image is shown on
\fref{fig:crystallite}.

\begin{figure}
 \centerline{\includegraphics[width=\normfig]{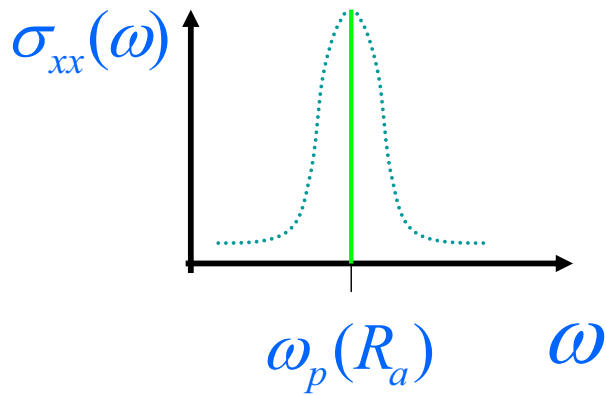}}
 \caption{In an image where the system is broken into crystallites of sizer $R_a$,
 one can imagine that each crystallite is experiencing a pinning force and is in an energy well.
 There is thus a natural pinning frequency, $\omega_p$ which is
 controlled by $R_a$. The conductivity has thus a peak at this frequency. Coupling between crystallites
 and distribution of pinning forces leads to a broadening of this pinning peak, very often chosen
 as lorentzian for lack of a better function.}
 \label{fig:condupin}
\end{figure}

If one assumes that all positional order is lost, this allows to
solve for the dynamics of the problem. In that case one can
consider that each crystallite is pinned independently. When one
tries to move it it sees a pinning potential that one can
approximate by an harmonic well. The shape of the well depends on
the size of the crystallite. If the system is submitted to an
electric field at frequency $\omega$, each crystallite will
respond as a particle in a well. Crudely speaking the minimal
frequency one can excite is (using $\omega = c q$ and $q_{\rm min}
\sim 1/R_a$) is
\begin{equation}
 \omega_{\rm pin} \propto \frac{c}{R_a}
\end{equation}
The perfect conductivity of the pure crystal, i.e. the
$\delta$-function peak that existed at zero frequency is now
pushed to the finite frequency $\omega_{\rm pin}$, as shown in
\fref{fig:condupin}.
If there is a finite magnetic field the peak at $\omega_c$ is
essentially unchanged and there is a second peak that appears at
$\omega_{\rm pin}$. Of course the approximation of independent
crystallites is not exactly true and the coupling between the
crystallites as well as the distribution of pinning strength leads
to a broadening of the peaks. The broadening is hard to determine
precisely since one does not know exactly the static solution, so
quite often a lorentzian broadening is assumed. This physical
image inspired from pioneering theories \cite{fukuyama_pinning}
used for charge density waves allows to compute, using some
approximations, the optical conductivity (see e.g.
\cite{giamarchi_quantum_pinning} for further details on that
point). This gives specific predictions, in particular the density
dependence of the pinning frequency would be from the above
mentioned approximations $\omega_p \propto n^{1/2}$.

\begin{figure}
 \centerline{\includegraphics[width=\normfig]{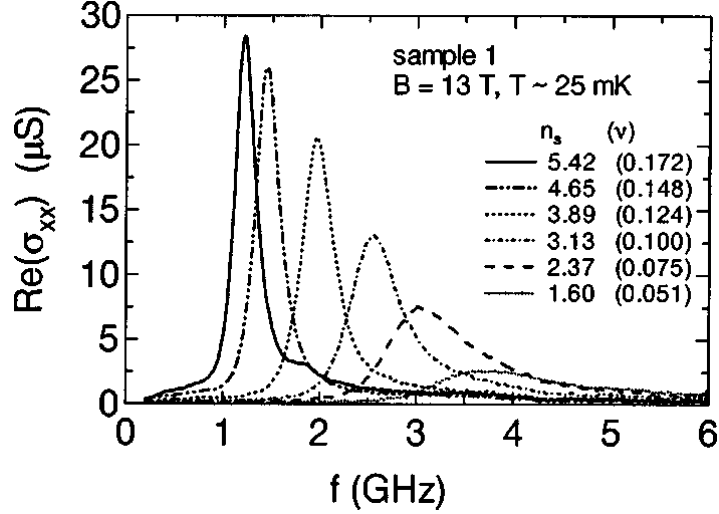}}
 \caption{Optical conductivity for various densities for a 2DEG
 under strong magnetic field. The peak at a characteristic
 frequency (pinning frequency) is an expected characteristics
 of a pinned crystalline structure. The density and magnetic field dependence
 of the pinning peak provide a stringent test for the existence of a pinned Wigner crystal
 phase (from \cite{li_conductivity_wigner_density}).}
 \label{fig:optics}
\end{figure}

\begin{figure}
 \centerline{\includegraphics[width=\normfig]{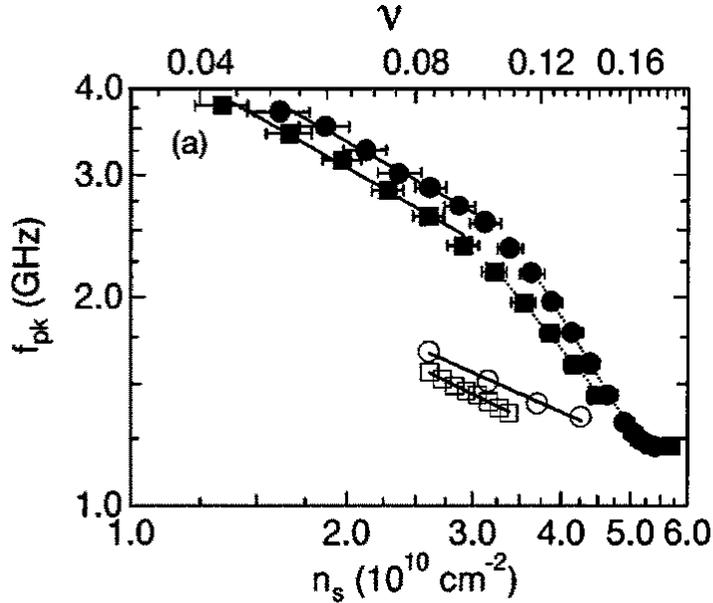}}
 \caption{Variation of the pinning frequency with the density for a
 2DEG under a strong magnetic field. The pinning frequency
 decreases with the density in all systems (different symbols), in
 contradiction with the naive calculations based on the physical
 image shown in \fref{fig:crystallite}. Such calculations
 would lead to $\omega_p \propto n^{1/2}$, in \emph{qualitative} disagreement
 with experiments. (From \cite{li_conductivity_wigner_density})}
 \label{fig:pinfreq}
\end{figure}

Measuring the optical conductivity is thus a very powerful way to
check for the presence of a pinned Wigner crystal. An example of
data is shown in \fref{fig:optics}.
One clearly observes a pinning peak in the optical conductivity.
However there are many problems. First the broadening is far from
being simply lorentzian as is obvious when simply looking at the
data. A much more serious problem comes from the density
dependence of the pinning frequency that can be extracted from the
above data and is shown in \fref{fig:pinfreq}.
When one compares the experiments, with these theoretical
predictions the agreement is {\it qualitatively} wrong. The
predicted result would be totally opposite to the data which shows
a {\it decrease} of the pinning frequency with the density. Such
important problems when one tries to compare with the data could
cast serious doubts on the interpretation of the insulating phase
in terms of a Wigner crystal and quite naturally other
interpretations for this phase have been proposed
\cite{zhang_hall_insulator}.

In fact, the reason of this discrepancy is that the
phenomenological image of nearly independent crystallites is
simply plain wrong. In order to be able to compute the transport
properties we thus need sharper tools to solve the problem of
pinned quantum crystals. These are the tools I now present.

\section{Methods} \label{sec:methods}

Let us see which tools of our theoretical toolbox one can use to
tackle a disordered problem. If we put a random term in our
Hamiltonian, then every observable depends on the specific
realization of the random potential. What is needed is then to
take an average of these observables with respect of the
probability distribution of the disorder to get the average
response of the system. This is of course a theoretical trick. A
real experimental system has usually only one realization of the
disorder, and the self-averaging comes from the fact that the
system is large enough so that little pieces of the system more or
less behave as independent sub-systems. Solving the problem for a
given specific realization of the disorder and averaging
afterwards is of course a totally impossible program. In addition
of being impossible it is in general stupid, since a given
realization of the disorder breaks many symmetries of the system.
Since a given realization of the disorder is not invariant by
translation, all correlation functions depend now on both
coordinates $x$ and $x'$ at which they are computed and not simply
on the difference $x-x'$ as in a translationally invariant system.
On the other hand the \emph{averaged} correlation function is
invariant by translation so it is much simpler. Of course averages
can be done order by order in a perturbation theory, but if one
wants to go beyond perturbation one wants a method to average over
disorder \emph{from the start}. Unfortunately many techniques,
that were useful for the non interacting systems (such as
supersymmetry \cite{efetov_supersym_revue}) fail when interactions
are included. I present here a quite general method that still
works and is known as the replica method. The other useful method
is the dynamical method (so called Keldysh technique) but it is of
a more complex use. For a recent review on dynamical methods see
\cite{kamenev_keldysh_review}. This method is specially important
if one wants to do out of equilibrium dynamics in quantum systems
since this is the only one that can deal with this question.

Let me assume that we want to compute the average value of some
observable $O$ for a system with a random potential $V$. The
average value can be written as a functional integral over the
configurations of the system as
\begin{equation} \label{eq:integ}
 \langle O \rangle_V = \frac{\int \DD\phi O(\phi) e^{-S_V(\phi)}}{\int \DD\phi
 e^{-S_V(\phi)}}
\end{equation}
where $S_V(\phi)$ is the action of the system for a given
realization of the random potential $V$. Of course $\langle O
\rangle_V$ depends on $V$ itself, so we have to average over all
realizations of $V$. If we assume that the disorder has a
probability distribution $p(V)$ the average over disorder is
\begin{equation} \label{eq:degeleq}
 \overline{\langle O \rangle}= \frac{\int \DD V p(V)
 \langle O \rangle_V}{\int \DD V p(V)}
\end{equation}
In general $S_V(\phi)$ is linear in $V$, something like (this is
indeed the case in the disordered crystal problem)
\begin{equation}
 S_V = S_0(\phi) + \int dx d\tau V(x) A(\phi(x,\tau))
\end{equation}
Note that for a quantum problem the disorder is time independent.
For the disorder one takes in general a gaussian disorder. This is
very often justified by the central limit theorem. For example
using a distribution
\begin{equation} \label{eq:disdisor}
 p(V) = e^{- \frac1{2D}\int dx V(x)^2} = e^{-\frac1{2D\Omega}\sum_q V^*_q V_q}
\end{equation}
corresponds to the average
\begin{equation}
 \overline{V(x)V(x')} = D \delta(x-x')
\end{equation}
With the distribution (\ref{eq:disdisor}) it would be very easy to
perform the average (\ref{eq:degeleq}) if it were not for the
denominator in (\ref{eq:integ}). Indeed in the absence of such
denominator one has
\begin{eqnarray} \label{eq:averagewrong}
 \frac{1}{\int \DD V e^{- \frac1{2D}\int dx V(x)^2}}\int \DD V e^{- \frac1{2D}\int dx V(x)^2}
 e^{-\int dx \int d\tau V(x) A(\phi(x,\tau))} &=&
 \nonumber \\
 e^{\frac{D}2 \int dx\int d\tau\int d\tau' A(\phi(x,\tau))A(\phi(x,\tau'))}
\end{eqnarray}
One would end up with an effective action where the disorder has
been eliminated and has given after average an interaction term so
the action would be
\begin{equation}
 S_{\rm eff} = S_0(\phi) - \frac{D}2\int dx\int d\tau\int d\tau' A(\phi(x,\tau))A(\phi(x,\tau'))
\end{equation}
we could then treat this problem using our favorite method since
it would not be more complicated than the type of problems that we
already encountered in this book.

The idea of the replica method \cite{edwards_replica} is thus to
get rid of the denominator and to transform it into a numerator
(see the lecture notes by I. V. Lerner in this volume, for the use
of replicas for calculation of the partition function). As with
any really great method the idea is very simple. One can rewrite
\begin{eqnarray}
 \frac1{\int \DD\phi e^{-S_V(\phi)}} &=& \left[\int \DD\phi
 e^{-S_V(\phi)}\right]^{n-1}
\end{eqnarray}
with $n=0$. If we forget $n=0$ for a moment and consider $n$ as a
positive integer $n=2,3,4,\ldots$, then
\begin{eqnarray}
 \left[\int \DD\phi e^{-S_V(\phi)}\right]^{n-1} &=&
 \left[\int \DD\phi_2 e^{-S_V(\phi_2)}\right]
 \cdots
 \left[\int \DD\phi_n e^{-S_V(\phi_n)}\right]
\end{eqnarray}
where we have introduced the fields $\phi_2$, $\phi_3$ etc. The
denominator can thus be rewritten as the product of $n-1$ copies.
The average (\ref{eq:integ}) can thus be rewritten
\begin{eqnarray} \label{eq:integrep}
 \langle O \rangle_V &=&
 \left[\int \DD\phi_1 O(\phi_1) e^{-S_V(\phi_1)}\right]
 \left[\int \DD\phi_2 e^{-S_V(\phi_2)}\right]
 \nonumber\\
 & & \left[\int \DD\phi_3 e^{-S_V(\phi_3)}\right] \cdots
 \left[\int \DD\phi_n e^{-S_V(\phi_n)}\right]
 \nonumber\\
 &=& \int \DD\phi_1\DD\phi_2\ldots\DD\phi_n O(\phi_1)
     e^{-\sum_{a=1}^n S_V(\phi_a)}
\end{eqnarray}
There is no denominator anymore. The price to pay is the
introduction of $n$ copies of the system. Of course one would only
recover (\ref{eq:integ}) if one can take the limit $n\to 0$ at the
end. Before averaging over disorder in (\ref{eq:integrep}) all
replicas (copies) are independent. Since there is no denominator
in (\ref{eq:integrep}) one can do the average over disorder in the
manner described above. One thus finds
\begin{eqnarray}
 \overline{\langle O \rangle}
 &=& \int \DD\phi_1\DD\phi_2\ldots\DD\phi_n O(\phi_1)
     e^{-S_{\rm eff}}
\end{eqnarray}
where the effective action is now
\begin{equation} \label{eq:effacrep}
 S_{\rm eff} = \sum_{a=1}^n S_0(\phi_a) -
 \frac{D}2\sum_{a=1,b=1}^n \int dx\int d\tau\int d\tau' A(\phi_a(x,\tau))A(\phi_b(x,\tau'))
\end{equation}
This is nearly the same form except that now one has $n$ fields
and the interaction couples all fields together. We have thus
traded a disordered system with only one field for a clean
interacting problem of $n$ coupled fields. Of course the second
one is more complicated due to the presence of the $n$ fields, but
as explained before we should be more equipped to tackle it. One
important difficulty is of course that we should obtain a good
enough solution for any $n$ to be able to make the analytic
continuation to $n \to 0$ at the end, since it is only in this
limit that one recovers the disordered solution. Taking this limit
is far from being obvious and contains hidden difficulties that I
will briefly allude to later in this section.

If we use this method we end up with an action $S = \Sel + \Sdis$
where $\Sel$ is the elastic part of the action and the disorder
part is
\begin{equation} \label{eq:disorderrep}
 \Sdis = - \rho_0^2 \Delta \sum_{a,b = 1}^n \sum_K \int dx d\tau d\tau'
 \cos(K(u_a(r,\tau)-u_b(r,\tau')))
\end{equation}

The term (\ref{eq:disorderrep}) is of course extremely complicated
to solve. To tackle such a term two paths are available. The first
one rests on the hope that the properties of the system are
dominated by small fluctuations around the minimum of one of the
cosine. One would thus hope that expanding the cosine is
relatively accurate. In fact things are more complicated because
of the presence of replicas. The second possible way is to tackle
the problem as a standard critical problem and try to find
renormalization equations for the disorder term. Here again the
situation is more complicated than one would naively think. I will
examine these two methods in the following.

\subsection{Variational method}

\begin{figure}
 \centerline{\includegraphics[width=\normfig]{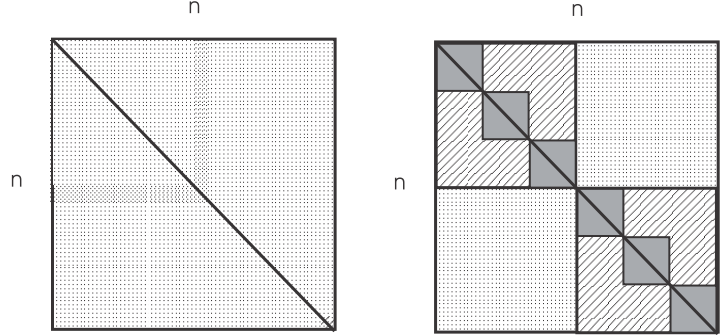}}
 \caption{(left) Replica symmetric structure. The matrix $G_{ab}$ has one value on the
 diagonal $\tilde G$ and one value away from the diagonal $G$. (right) Replica symmetry breaking
 structure. The example of a two step RSB is shown. The matrix $G_{ab}$ is a hierarchical matrix.
 There is one value $\tilde G$ on the diagonal and several values (here three) away from the diagonal.
 In the limit $n\to 0$ the value away from the diagonal becomes a function $G(u)$ labelled by an index $u\in [0,1]$,
 giving its position in the hierarchical structure. In this example the function $G(u)$ takes three values
 with two breakpoints (hence the name two step RSB).}
 \label{fig:rsvsrsb}
\end{figure}

The first method is to try to reduce this problem to the ``best''
quadratic problem approximating $\Sel + \Sdis$. The most brutal
way would be to simply expand the cosine, but as is well known for
sine-Gordon problems this is very often a poor approximation. Here
is would be a complete catastrophe for reasons that will become
clear later. An improved way is to perform a self consistent
calculation. The simplest way to implement such a scheme it
through Feynman's variational approach \cite{feynman_statmech}.
The method has been extended to deal with replicas in
\cite{mezard_variational_replica}. Let me define a trial action
$S_0$. Then the free energy $F$ of the system verifies
\begin{equation} \label{eq:varbas}
 F \leq \Fvar = F_0 + \langle S - S_0 \rangle_0
\end{equation}
where $\langle\rangle_0$ denotes an average taken with the action
$S_0$. If $S_0$ depends on some variational parameter one can hope
that the ``best'' way (i.e. the one giving a physics as close as
possible to the one of the true action $S$) of choosing this
parameter is to minimize $\Fvar$. Let us take for $S_0$ the most
general quadratic action
\begin{equation}
 S_0 = \frac{1}{\beta\Omega} \sum_{ab}\sum_{q,\omega_n} G^{-1}_{ab}(q,\omega_n)
 u_a^*(q,\omega_n) u_b(q,\omega_n)
\end{equation}
The choice of a quadratic action for $S_0$ has two motivations:
(i) quadratic actions are the only ones with which one is able to
compute something; (ii) on a physical basis one can expect the
properties of the system to be dominated by the small oscillations
around one of the minima of the cosine. We will have to come back
to this assumption. In  doing so one implicitly neglects
excitations that can take the system from one minimum of the
cosine to the other. For the sine-Gordon model these excitations
are known as solitons \cite{rajaraman_instanton}. For the pure
problem this method is equivalent to the well known self
consistent harmonic approximation, and gives excellent results.
The disordered case is more involved
\cite{giamarchi_columnar_variat,giamarchi_quantum_pinning}. In
particular special case should be paid to the mode $\omega_n = 0$
since the disorder is time independent. I will only sketch the
main lines here. The full Green's functions are here the
variational parameters. They are determined by
\begin{equation}
 \frac{\partial \Fvar}{\partial G_{ab}(q,\omega_n)} = 0
\end{equation}
Using (\ref{eq:varbas}) leads to nasty integral equations, roughly
of the form
\begin{equation} \label{eq:oftheform}
 G^{-1}_{ab}(q,\omega_n) = (\omega_n^2 + q^2)\delta_{ab} + \int
 d\tau [1 - \cos(\omega_n \tau)] e^{-
 \frac{1}{\beta\Omega}\sum_{q,\nu_n}G_{ab}(q,\nu_n) \cos(\nu_n
 \tau)} \ldots
\end{equation}
One has to remember that the limit $n \to 0$ should be taken at
the end so $G_{ab}$ is in fact a $0\times0$ matrix. In order to
invert this matrix to solve the equations (\ref{eq:oftheform}) one
needs to know what is the structure of such matrices. In the
original action, $\Sel$ is purely diagonal in replicas, whereas
$\Sdis$ couples each replica with another and is thus independent
of the value of the indices $a,b$. This suggests that a good
structure for $G_{ab}$ is to take one (for each value of $q$ and
$\omega_n$ of course) value on the diagonal, and one \emph{unique}
value away from the diagonal
\begin{equation} \label{eq:rs}
 G_{ab} = [\tilde G - G] \delta_{ab} + G
\end{equation}
Such a matrix structure is shown in \fref{fig:rsvsrsb}.
This is the \emph{natural} choice, and is called replica symmetric
since it respects the permutation symmetry between the replicas.
The matrix depends in fact on only two numbers: the value on the
diagonal $\tilde G$ and the off diagonal value $G$. It is easy to
see that such matrices can be inverted and have a well defined
limit when $n\to 0$. Calling $\tilde G^{-1}$ and $G^{-1}$ the
diagonal and off diagonal value of the matrix $G^{-1}_{ab}$ and
the connected part $G_c = \sum_b G_{ab}$ one has the inversion
formulas (in the limit $n\to 0$)
\begin{equation}
 G_c = \frac{1}{G_c^{-1}} \quad,\quad G = - \frac{G^{-1}}{(G_c^{-1})^2}
\end{equation}
This allows to close the integral equations (\ref{eq:oftheform})
and in principle to solve them. Unfortunately, it is easy to see
that (\ref{eq:rs}) cannot be right. This can be seen if one
brutally expand the cosine. This indeed corresponds to a RS
structure and gives ``directly'' $S_0$ (this approximation
corresponds to the variational approach if one \emph{assumes} the
RS structure and that the disorder $\Delta$ is large). In that
case only the $\omega_n = 0$ frequency appears due to the double
integration on $\tau$ and $\tau'$. Since the action is quadratic
this means that the disorder would only affect the mode $\omega_n
= 0$, and thus would be unable to change the current. This is
obviously completely opposite to the idea of pinning, so such an
approximation misses all this physics. The answer to this paradox
resides in the assumption of the replica symmetric structure for
the matrix $G_{ab}$. Indeed as is now well known for spin glasses,
such a saddle point might be unstable \cite{dealmeida_thouless}.
In that case one should look for another structure for $0\times0$
matrices. This structure was discovered by Parisi, and breaks the
replica symmetry \cite{mezard_book}. The matrix is now a
hierarchical matrix. There is not a single element off diagonal
but many values defined in a hierarchical way. I will not enter in
the structure of the theory, and refer the reader to
\cite{giamarchi_book_young,giamarchi_quantum_pinning} for a
discussion. A pictorial representation of the matrix is shown in
\fref{fig:rsvsrsb}. Such matrices can also be inverted in the
$n\to 0$ limit. This structure is called replica symmetry
breaking. In this limit the off diagonal element becomes a
function $G(v)$ of a continuous variable varying between zero and
one. If $G(v)$ is a continuous function one speaks of continuous
RSB, whereas if $G(v)$ has step(s) one talks about $n$-step RSB
($n$ being the number of steps). I will not dwell too much on the
physical interpretation and refer the reader to
\cite{giamarchi_quantum_pinning}. As is well known for spin
glasses, the occurrence of RSB signals glassiness, and the
presence of metastable states. Since there is one matrix
$G_{ab}(q,\omega_n)$ for each value of $q$ and $\omega_n$, the
replica structure depends a priori on $q$ and $\omega_n$. For
quantum problems since the disorder is time independent one can
show \cite{giamarchi_columnar_variat} that one has off diagonal
replica terms \emph{only} for the mode $\omega_n = 0$. The
structure is replica symmetric for $q
> 1/R_c$ and RSB for $q < 1/R_c$. This confirms the physical
picture that pinning and metastability appear above the Larkin
length $R_c$.

\subsection{Functional renormalization group}

\begin{figure}
 \centerline{\includegraphics[width=\normfig]{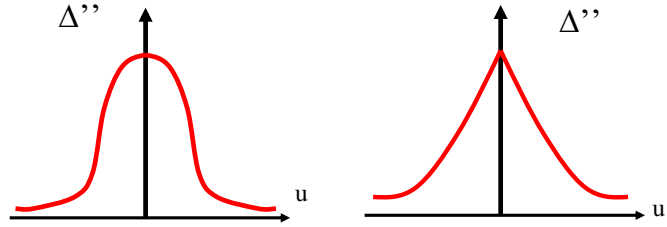}}
 \caption{In the FRG approach one has to renormalize the full correlator of the disorder.
 One starts with an analytic function (left). For crystals this function is essentially a cosine.
 After a certain lengthscale, corresponding to the Larkin length, the function develops a nonanaliticity
 (cusp) at the origin (right). This nonanaliticity is the signature in this method of the glassy properties
 of the system.}
 \label{fig:cusp}
\end{figure}

RSB might look like magic\footnote{This is metaphysical question,
since the only point is whether a method works or not!}, so let us
try to solve our disordered problem by a more conventional
renormalization approach. Doing it on quantum problems is tedious
so I will explain the method on a classical system. One simply
takes $\hbar = 0$ in the quantum action which removes all time
dependence in time of $u(r,\tau)$. The system is thus described by
the classical action
\begin{eqnarray} \label{eq:acdep}
 S &=&  \frac{c}{2T} \sum_a\int d^dr (\nabla u_a(r))^2
   -\frac{\Delta}{T^2}\sum_{ab}\int d^dr \cos(K(u_a(r)-u_b(r)))
\end{eqnarray}
where I have retained only a single harmonic in the disorder term
for simplicity. To implement an RG procedure one would decompose
$u_a(r)$ into fast and slow modes. If for example $\Lambda$ is a
cutoff in $q$ space and $\Lambda'$ a smaller cutoff, one defines
($\Omega$ is the volume of the system)
\begin{eqnarray}
 u_a(r) & =& u_a^<(r) +  u_a^>(r) \nonumber \\
  &=& \frac{1}{\Omega} \sum_{q < \Lambda'} u_a(q) e^{i q r}
     + \frac{1}{\Omega} \sum_{\Lambda' < q < \Lambda} u_a(q) e^{i q r}
\end{eqnarray}
One then integrates over the fast modes ($\Lambda' < q < \Lambda$)
to get an effective action with new coupling constants for the
slow modes. Since the cutoff has changed it is then convenient to
rescale space by $q = q' \Lambda'/\Lambda$ or $r = r'
\Lambda/\Lambda'$, such that the new space variable $q'$ has again
a cutoff $\Lambda$. One can then iterate this procedure. Even
without performing the integration over the fast modes let me look
at the bare dimensions of the operators (tree level). Let us
assume that $u$ scales as $u \sim L^\zeta$. One sees directly that
periodic systems are peculiar. Because the disorder term is a
cosine, with a periodicity which corresponds to the lattice
spacing, this periodicity cannot be changed in the
renormalization. This imposes $\zeta = 0$
\cite{giamarchi_vortex_short,giamarchi_vortex_long}. Let me
parametrize the cutoff by $\Lambda(l) = \Lambda_0 e^{-l}$, where
$\Lambda_0$ is the initial cutoff. In order to keep the elastic
part of the action invariant, one should thus have
\begin{equation} \label{eq:scaltemp}
 \frac{1}{T(l+dl)} = \frac{1}{T(l)}
 \left(\frac{\Lambda(l)}{\Lambda(l+dl)}\right)^{d-2}
\end{equation}
the $d-2$ comes from the integration $d^dr$ and the $\nabla^2$.
This gives the scaling dimension for the temperature
\begin{equation}
 \frac{d T(l)}{dl} = T(l) (2-d)
\end{equation}
We thus see that for $d > 2$ the temperature scales to zero. This
is in agreement with the fact that the competition between
disorder and elasticity is an \emph{energy} problem, where the
temperature plays little role. Let us look at the renormalization
of the disorder term. Performing the same rescaling of $d^dr$ and
using (\ref{eq:scaltemp}) for the temperature gives
\begin{equation}
 \Delta(l+dl) = \Delta(l) \left(\frac{\Lambda(l+dl)}{\Lambda(l)}\right)^{d-4}
\end{equation}
giving the equation
\begin{equation}
 \frac{d \Delta(l)}{dl} = \Delta(l) (4-d)
\end{equation}
We thus see that the disorder is relevant below $d=4$. Note that
around $d = 4$ the temperature is indeed irrelevant. One could
thus naively think that this problem is quite similar to the
standard $4-\epsilon$ renormalization in critical phenomena. There
is however an important difference. Let us compare the present
problem with a standard (non disordered problem) in the standard
$\phi^4$ theory. Let me denote the action for this problem
\begin{equation} \label{eq:phifour}
 \frac{c}{2T} \int d^dr (\nabla\phi)^2 + \frac{\lambda}{T} \int
 d^dr \phi^4 + \frac{\mu}{T} \int
 d^dr \phi^6
\end{equation}
In that case and contrarily to the disorder case both the elastic
term and the ``potential'' terms have the \emph{same} power of the
temperature. This is of course the signature that a normal phase
transition results from a competition energy-entropy. The extra
factor of $1/T$ in front of the cosine in (\ref{eq:acdep}) is the
way for the replicated action to indicate that \emph{before}
averaging one had to compare the two energies of the elastic term
and of the disorder term. For (\ref{eq:phifour}) one absorbs the
factor $e^{-(d-2)dl}$ coming from the spatial rescaling in the
rescaling of $u$ and one leaves the temperature \emph{invariant}.
This traduces the fact that a problem at $T = T_c$ stays at $T_c$.
The rescaling for $u$ thus becomes
\begin{equation}
 \phi^2(l+dl) = \phi^2(l) e^{(d-2)dl}
\end{equation}
This gives for the renormalization of the coupling constants
\begin{eqnarray}
 \frac{d \lambda}{dl} &=& \lambda(4-d) \nonumber \\
 \frac{d \mu}{dl} &=& \mu(6-2d)
\end{eqnarray}
Thus all terms in the potential that have high (i.e. larger than
$4$) powers of $u$ are \emph{irrelevant}. Thus regardless of the
precise form of the potential, one can always expand in power
series and the only important term is the $\phi^4$ term. The
situation is totally different for the disordered system
\cite{fisher_frg_1}. In that case $u$ does not scale. It means
that \emph{all} terms in a power series expansion of the potential
have the same scaling dimension and are \emph{equally} relevant
around four dimensions. Thus instead of having a single RG
equation for a single parameter $\lambda$ one has now an
\emph{infinity} of (a priori coupled) RG equations for all the
coefficients of the terms in the power series expansion of the
potential. In other words for the disorder problem one has now to
renormalize the \emph{whole function} corresponding to the
potential term. This means that one has to start with a potential
term
\begin{equation}
 \Sdis = -\frac{1}{T^2} \sum_{ab} \int d^dr \Delta(u_a(r)-u_b(r))
\end{equation}
where $\Delta(z)$ is a function whose initial ($l=0$) value is
$\Delta_{l=0}(z) = \cos(K z)$ but which will change form in the
course of the renormalization. Because one has to keep the whole
function under renormalization this procedure is known as
functional renormalization group. I will not derive the equations
to the next order (a good description of the derivation can be
found in an appendix of \cite{balents_frg_largen}) but just give
the result (for a classical crystal)
\cite{giamarchi_vortex_short,giamarchi_vortex_long}
\begin{equation} \label{eq:basfrg}
 \frac{d\Delta(z)}{dl} = \epsilon \Delta(z) + \frac12\Delta''(z)^2 -
 \Delta''(z=0)\Delta''(z)
\end{equation}
Needless to say this is much more complicated to solve than if one
had a single parameter. But in addition there is an important
feature of the solution. Because we eliminate modes $u_a$ becomes
smoother and smoother thus in a sense $u_a-u_b$ becomes smaller
and smaller under renormalization. Thus if the function $\Delta$
stayed analytic one could expand it in powers of $u$ to get
\begin{equation}
 \Delta(u_a-u_b) = \Delta(0) + \frac12\Delta''(0)(u_a-u_b)^2
 + \frac{1}{4!}\Delta^{(4)}(0)(u_a-u_b)^4 + \ldots
\end{equation}
and one would be asymptotically dominated by the $(u_a-u_b)^2$
term. This would give back a RS solution, and in fact exactly
Larkin's result. Since we know that this is incorrect beyond a
certain lengthscale, something has to go wrong with the above
expansion. The only way to escape this is if $\Delta^{(4)}(0)$
becomes very large (in fact infinite) so that the above expansion
breaks down and that higher power continue to play a role. Thus to
get the glassy physics the function $\Delta(z)$ should become
\emph{nonanalytic} beyond a certain lengthscale. One can check
that this is indeed the case. The equation for $\Delta^{(4)}(0)$
is from (\ref{eq:basfrg})
\begin{equation}
 \frac{d\Delta^{(4)}(0)}{dl} = \epsilon \Delta^{(4)}(0) + 3
 [\Delta^{(4)}(0)]^2
\end{equation}
which indeed explodes at the lengthscale
\begin{equation}
 e^l = \left(\frac{\epsilon}{3 \Delta_0}\right)^{1/\epsilon}
\end{equation}
This corresponds indeed to the Larkin scale (compare with
(\ref{eq:larkinlength})). Beyond this lengthscale, although the
function itself remains finite, it develop a non-analyticity. If
one looks at the second derivative of $\Delta$ it develops a
\emph{cusp}, as shown on \fref{fig:cusp}.
I will stop there on the FRG method since this would take us way
beyond the level of these notes. For more on the FRG and further
references see
\cite{giamarchi_book_young,nattermann_vortex_review,ledoussal_frg_twoloops,wiese_frg_review}

\section{Bragg glass and disordered Wigner crystal} \label{sec:quantit}

\subsection{Elastic model and experiments}

Let me now examine what are the results for the disordered Wigner
crystal. I will not reproduce the calculations here that can be
found in
\cite{giamarchi_columnar_variat,giamarchi_quantum_pinning} for
quantum crystals in general and in
\cite{chitra_wigner_hall,chitra_wigner_long} for the specific case
of the Wigner crystal.

\begin{figure}
 \centerline{\includegraphics[angle=-90,width=\normfig]{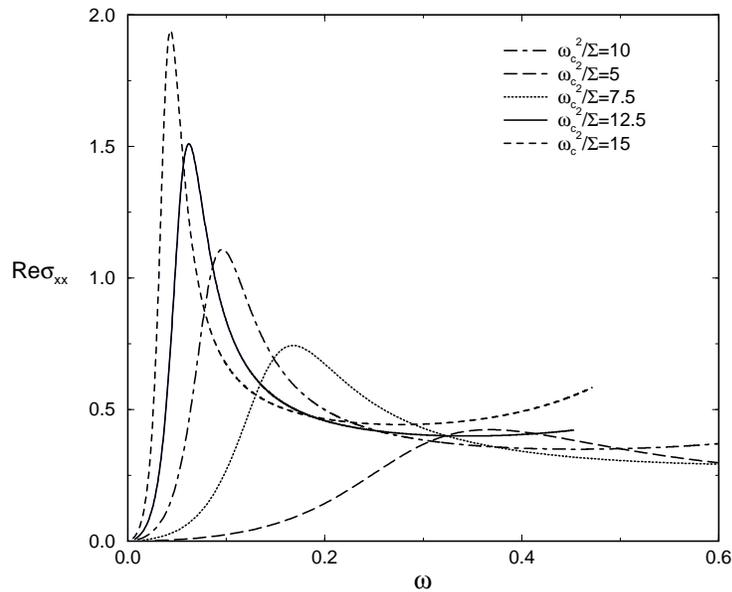}}
 \caption{An example of the conductivity of the Wigner glass as computed by the variational approach.
 Note the non lorentzian bradening of the peak and the magnetic
 field dependence of the pinning frequency (from \cite{chitra_wigner_long}).}
 \label{fig:conduvar}
\end{figure}

Using the variational method on (\ref{eq:ham}) and (\ref{eq:sdis})
one finds that the most stable solution is an RSB one. This is
suggestive that the pinned system indeed has glassy properties due
to the pinning. I will thus use loosely the term Wigner glass in
the following. Second, since we know have from the
displacement-displacement propagator, one can compute the
conductivity. An example of the results is shown in
\fref{fig:conduvar}.
At first sight the agreement with the experimental data is much
better (see \cite{chitra_wigner_long} for a full discussion). Note
that the broadening is non lorentzian in agreement with the data.
More importantly one finds a decrease of the pinning frequency
with the density as $\omega_p \propto n^{-3/2}$, as well as a good
magnetic field dependence of the pinning frequency. This very good
agreement gives a good confirmation that the insulating phase in
the 2DEG under strong magnetic field is indeed a Wigner crystal
collectively pinned by impurities.

What is the reason of this difference with the naive answers
explained in \sref{sec:wisdom}? An inspection of the variational
solution \cite{chitra_wigner_hall,chitra_wigner_long} shows that
in the full solution the pinning frequency is \emph{not}
controlled by $R_a$ as in the naive answer but by $R_c$ (in fact
the formula is more complicated). This is a very satisfactory and
physical result. Indeed as we discussed, for a classical crystal
$R_c$ is the length that controls the pinning force. Although is
it of course impossible to directly extrapolate it is quite
reasonable to obtain that the pinning frequency, that is another
manifestation of pinning, is also controlled by this lengthscale.
This difference is crucial since, as in the Wigner crystal $l_c$
and $a$ are quite different $R_c$ and $R_a$ corresponds to quite
different lengthscales and have in general quite different
dependence in the various parameters. Since $R_c$ depends on the
size of the particle $l_c$, this gives, for example for the case
of strong magnetic field for which $l_c$ is just the cyclotron
orbit, additional magnetic field dependence to the pinning
frequency. This is to be contrasted with charge density waves for
which $l_c \sim a$ due to the nearly sinusoidal density modulation
and thus $R_c \sim R_a$. Thus borrowing directly the solutions
that have been developed \cite{fukuyama_pinning} for this case is
dangerous and gives part of the physics incorrectly.

I refer the reader to
\cite{chitra_wigner_hall,chitra_wigner_long}, for other physical
properties such as the Hall conductivity. I will just mention here
the compressibility. Defining the compressibility for a charged
system is a delicate question since one has to maintain the
neutrality of the system. Naively one relates the compressibility
to the density-density correlation function by
\begin{equation} \label{eq:compress}
\kappa = \lim_{q\to 0} \langle \rho(q,\omega_n=0)
\rho(-q,\omega_n=0) \rangle
\end{equation}
Let me look first at the pure system. The compressibility is
simply (only the longitudinal mode plays a role)
\begin{equation}
\kappa(q) = \lim_{q\to 0} \frac{q^2}{c_L(q)}
\end{equation}
If only short range interactions are present in the system the
longitudinal mode is a phonon-like mode $c_L(q) \propto q^2$ and
the one recovers a finite compressibility. On the other hand if
one has long range Coulomb interactions $c_L(q) \propto  |q|$ and
the compressibility becomes zero. This is simply due to the fact
that (\ref{eq:compress}) measures the density response to a change
of chemical potential while {\it keeping the neutralizing
background unchanged}. A charged system thus does not remain
neutral, hence the infinite compressibility. One has thus to
define the compressibility more precisely. Many derivations of the
compressibility use instead directly a derivative of the free
energy with respect to the number of particles. The free energy
can be computed for a neutral system for an arbitrary number of
particles which solves the above-mentioned problem. Unfortunately
very often the calculation is only possible in some sort of
approximate way such as an Hartree-Fock approximation. Here again
the link with the direct measurements of the compressibility is
not clear. Using such procedures, so called ``negative''
compressibilities are found for some range of the interactions,
for interacting electrons. Similarly, experiments measure such
negative ``compressibilities''
\cite{eisenstein_hall_compressibility,ilani_compressibility_2DEG}
(see \cite{abrahams_review_mit_2d} for further references and
discussion on this question).

\begin{figure}
 \centerline{\includegraphics[width=\smallfig]{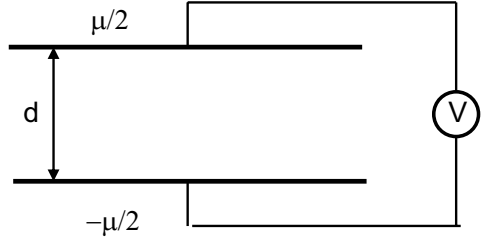}}
 \caption{Capacitance measurement, which gives access to the
 compressibility of the system. A voltage difference $\mu$ is
 applied to a capacitor. Here for simplicity the capacitor is made
 of two plates of the 2DEG.}
 \label{fig:capacitance}
\end{figure}

In order to make the physics of such negative compressibility more
transparent, I discuss now a very simple way to compute it
\cite{giamarchi_wigner_review}. It is in fact a direct calculation
of the quantity that is actually measured to determine the
``compressibility'', i.e. the capacitance of a system made by with
the 2DEG \cite{eisenstein_hall_compressibility}. For simplicity I
take here a capacitor formed of two identical systems, as shown in
\fref{fig:capacitance}.
Taking one system and one metallic plate would not change the
results in an essential way. The Hamiltonian of the system is thus
\begin{equation} \label{eq:hamdep}
 H_1^0 + H_2^0 + \sum_{(\alpha,\beta)=1,2} \int_{r,r'} \frac12 V(r-r')
 [\rho_\alpha(r) - \rho_0][\rho_\beta(r') - \rho_0]
 + \frac{\mu}2 \int_r [\rho_1(r) - \rho_2(r)]
\end{equation}
If one assumes that the system is neutral in the absence of $\mu$,
then the charge on one plate when a potential $\mu$ is applied is
\begin{equation} \label{eq:congen}
 \langle \rho_1 \rangle = \frac{\mu}2 [ \langle \rho_1 \rho_1 \rangle -
 \langle \rho_1 \rho_2 \rangle ]
\end{equation}
in linear response. (\ref{eq:congen}) give directly the
capacitance $\langle \rho_1 \rangle/\mu$.

One can use the general formula (\ref{eq:congen}) to compute the
capacitance for the Wigner crystal. One substitutes in
(\ref{eq:ham}) the density decomposition (\ref{eq:fourdens}). The
$\nabla u$ terms give directly the contribution of the long range
part of the Coulomb interaction
\begin{equation} \label{eq:lr}
 H_{\rm long-range} = \frac12 \rho_0^2 \sum_q \sum_{\alpha\beta
 =1,2}[V_{\alpha\beta}(q)u_L^\alpha(q)u_\beta(-q)]
\end{equation}
Since $V_{11}(q) \sim 1/q$, (\ref{eq:lr}) gives obviously the part
proportional to $q$ in the elastic coefficients for an isolated
plane. The higher harmonics give the regular part (i.e. the part
proportional to $q^2$ in the elastic coefficients. Such a way to
determine the coefficient is equivalent the calculation of the
coefficients in \cite{bonsall_elastic_wigner}.

Taking a pure system the Hamiltonian becomes (only the
$\omega_n=0$ term of the longitudinal part needs to be computed to
have the compressibility)
\begin{equation}
H =  \left(\begin{array}{c} u_L^1(q) \\ u_L^2(q) \end{array}
\right)
 =
 \left(\begin{array}{cc} c_L^{\rm SR}(q) +  \rho_0^2 q^2 V_{11}(q)& \rho_0^2 q^2 V_{12}(q) \\
                       \rho_0^2 q^2 V_{12}(q)   & c_L^{\rm SR}(q) +  \rho_0^2 q^2 V_{11}(q)
       \end{array} \right)
 \left(\begin{array}{c} u_L^1(-q) \\ u_L^2(-q) \end{array} \right)
\end{equation}
where $c_L^{\rm SR}(q)$ is the ``short range'' part of the elastic
coefficients. Using (\ref{eq:congen}) and the expression of the
density for small $q$ from (\ref{eq:fourdens}) $\rho_L(q) =
-\rho_0 q u_L(q)$ one gets for the capacitance $\frac1{C} =
\frac1{C_{\rm geom}} + \frac1{C_{\rm el}}$, where $C_{\rm geom} =
1/(4\pi d)$ is the standard geometrical capacitance and
\begin{equation} \label{eq:capa}
\frac1{C_{\rm el}} = \lim_{q\to 0} \frac{2 c_L^{\rm
SR}(q)}{\rho_0^2 q^2}
\end{equation}
(again the factor of $2$ comes from the fact that here I took two
identical plates). The electronic capacitance corresponds to the
propagator where {\it only} the short range part of the elastic
coefficients is kept. Using \cite{bonsall_elastic_wigner}
$c_L^{\rm SR}(q) = - \omega_0^2(0.18..)(a q)^2$, where $\omega_0 =
\frac{4\pi e^2}{\sqrt3 m a^3}$, one finds for the Wigner crystal a
``negative'' compressibility. The fact that a system of discrete
charges can lead to such effects has been noted before for
classical Wigner crystals (see e.g.
\cite{nguyen_wigner_biology_houches} and references therein).

The disorder can be considered easily. The variational method
discussed in \sref{sec:quantit} gives straightforwardly the
density-density correlation function which is simply related to
the $\langle u u \rangle$ correlation. Since it is a gaussian
approximation one can easily use the capacitance method shown
above. The capacitance is thus given by (\ref{eq:capa}) but where
one should use the propagator in the presence of disorder. It is
given by (within the variational approximation used
\cite{chitra_wigner_hall,chitra_wigner_long}) by:
\begin{equation} \label{eq:cdis}
 \langle u u \rangle = \frac{1}{\rho_m \omega_n^2 + c_L(q) + \Sigma(1-\delta_{n,0}) + I(\omega_n)}
\end{equation}
where $\Sigma$ and $I(\omega_n)$ are respectively  a constant and
a function related to the disorder verifying $I(0) = 0$. At
$\omega_n=0$ (\ref{eq:cdis}) leads to a compressibility in
identical to the one of the pure system, and thus {\it also}
``negative''. More on this point can be found in
\cite{giamarchi_wigner_review}.

\subsection{Defects} \label{sec:defects}

We thus see that our elastic description of an electronic glass is
a very good starting point to extract the physical properties
since we have now reliable analytic methods to tackle such
problems. However, in view of the commonly accepted point of view
that disorder would induce topological defects in a crystal (as
explained in \sref{sec:wisdom}) is it important to ascertain the
validity of the elastic description itself. In fact the creation
of defects by disorder has been grossly overestimated and the
elastic theory is in fact much more stable to the presence of
topological defects than initially anticipated following the naive
picture of \sref{sec:wisdom}.

\begin{figure}
 \centerline{\includegraphics[width=\verysmallfig]{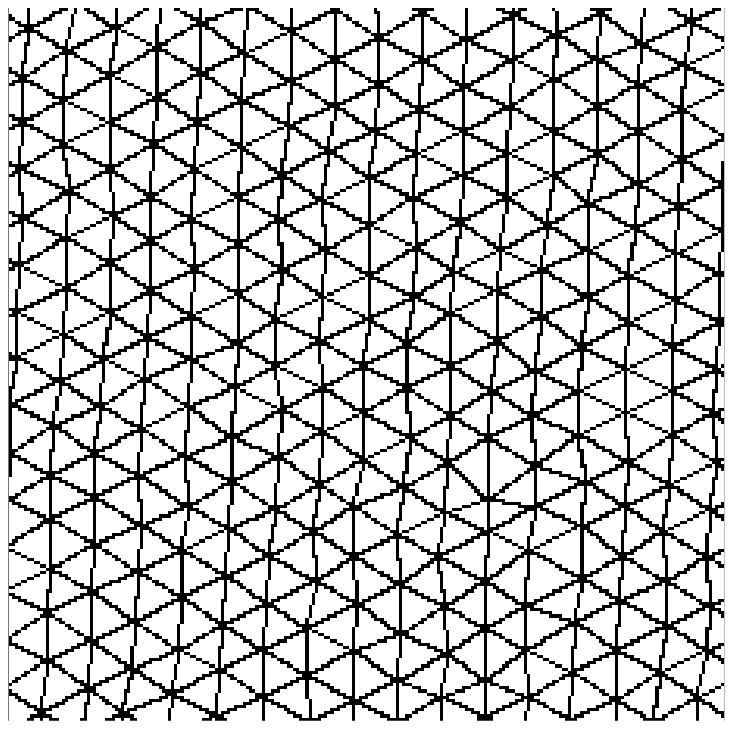}}
 \caption{In $d=3$ a classical crystal plus disorder is in a Bragg glass state (for weak disorder). Although the disorder
 destroys the perfect positional order of the crystal, it does it smoothly. The topological
 order of the crystal is preserved and there are no unpaired defects such as dislocations generated by the
 disorder. The system is thus nearly as ordered as a perfect crystal. In particular the relative displacements
 of two neighbors remain small, and thus such a system is totally described by an elastic theory.}
 \label{fig:imabragg}
\end{figure}

In $d=3$ it is now known that below a certain threshold of
disorder {\it no} topological defects can be induced by the
disorder. The disordered elastic system is in a Bragg glass state
\cite{giamarchi_vortex_long} with a quasi long range positional
order, much more ordered than the incorrect image of
\fref{fig:crystallite} suggests (see e.g.
\cite{giamarchi_vortex_review,nattermann_vortex_review} for a
discussion and references on this point). The distance between
unpaired defects is thus infinite in this case. Though although
the crystal is distorted by the disorder and the perfect
positional order is lost, the deformation is smooth and the
\emph{topological} structure of the crystal is preserved. The
displacements of two neighbors remain small even if large
distortions can accumulate at large lengthscales. The elastic
description is thus perfectly accurate at all lengthscales in that
case. An image of the Bragg glass is indicated in
\fref{fig:imabragg}.

\begin{figure}
 \centerline{\includegraphics[width=\normfig]{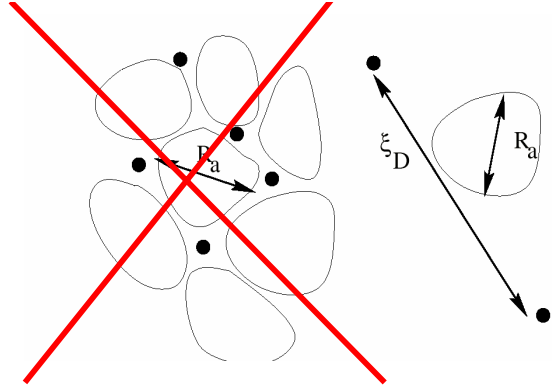}}
 \caption{In $d=2$ for a classical crystal, defects are generated by disorder at large lengthscales, so an elastic
 description is not valid beyond the distance $\xi_d$ between these topological defects. However for weak
 disorder $\xi_D$ is much larger than the lengthscale $R_a$ (right). This is in marked contrast with the incorrect
 naive picture of \fref{fig:crystallite} (left), where both length would be
 of the same order. It means that an elastic theory can reliably
 be used to describe the effects of disorder below $\xi_D$ (from \cite{ledoussal_dislocations_2d}).}
 \label{fig:def2d}
\end{figure}

\begin{figure}
 \centerline{\includegraphics[width=\normfig]{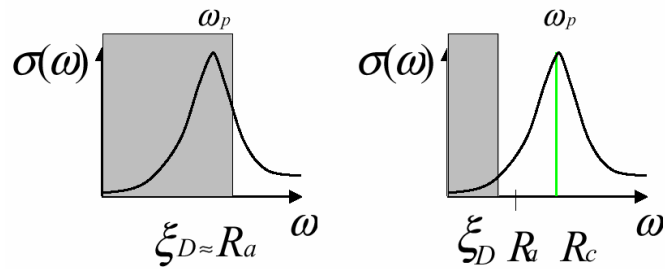}}
 \caption{(a) If dislocations  occurred at scale $R_a$ and the
 pinning frequency was controlled by $R_a$, as was naively
 believed, the elastic theory is incapable of giving any reliable
 information on the pinning peak. It would be necessary to include
 dislocations from the start. (b) As was discussed in the text,
 dislocations occur in fact at $\xi_D \gg R_a$ and the pinning peak
 depends on $R_c \ll R_a$. Thus the pinning peak is given
 quantitatively by a purely elastic theory. It is necessary to take
 into account topological defects such as dislocations only at much
 lower frequencies, and in particular if one wants to obtain
 reliable results for the d.c. transport (from
 \cite{chitra_wigner_long}).}
 \label{fig:disloc}
\end{figure}

In $d=2$ the situation is marginal, and defects do appear in the
ground state, but at distance $\xi_D$ much larger (for weak
disorder) than the lengthscale $R_a$ and {\it not} at that
lengthscale
\cite{giamarchi_vortex_long,ledoussal_dislocations_2d}, as shown
in \fref{fig:def2d}.
Since the ratio $\xi_D/R_a$ can become arbitrarily large for weak
disorder, it means that all important effects of disorder (they
occur at the lengthscales $R_c$ and $R_a$) can be reliably
described by an elastic theory. Since the pinning peak is
controlled by the length $R_c \ll R_a \ll \xi_D$ it means that
dislocations will spoil the results of the elastic theory for the
optical conductivity only \emph{well below the pinning peak}, as
shown on \fref{fig:disloc}. This implies that the theory is a {\it
reliable tool} to compute the characteristics of the peak and
above, and thus most of the a.c. transport.
Thus optical conductivity measurements are an accurate way of
probing the Wigner glass nature of the system, since one has a
reliable theory to compute the characteristics of the pinning peak
in a Wigner glass phase. This allows for a detailed testing, as I
showed above for the case of the 2DEG under strong magnetic field.
For the systems in the absence of a magnetic field shown in
\fref{fig:kravchenko}, a measurement of the optical conductivity
could thus provide an unambiguous answer on the presence of a
Wigner glass in the insulating phase. Such an experiment remains
to be performed.

On the contrary this shows that to know d.c. conductivity it is
\emph{necessary} to take topological defects of the crystal into
account (vacancies, interstitials, dislocations etc.). This makes
the calculation of the d.c. conductivity much more difficult since
at the moment there is no theory able to take into account
topological defects, elasticity and disorder at the same time. It
is thus impossible to make specific predictions specific to the
Wigner crystal and that could be quantitatively checked. The only
case where such a calculation of d.c. conductivity can be done
reliably is the case of one dimension (Luttinger liquids)
\cite{nattermann_temperature_luttinger}.

\section{Further steps and unsolved questions} \label{sec:conclusion}

In these short notes I have discussed some of the physics of
quantum particles in the presence of strong repulsion. In that
case the physics changes drastically from the standard ``wave''
description of indiscernable particles. The particles localize to
form a quantum crystal. Examples of such systems are the Wigner
crystal and Luttinger liquids. In such a quantum crystal, quantum
effects are still important but particles become essentially
discernable by their position, making the problem much simpler to
tackle. The effects of disorder in such a disordered crystal are
also quite different than on noninteracting electrons, and can be
viewed as the pinning of the quantum crystal on impurities, in a
similar way than for a classical crystal.

I have briefly described the methods used to tackle such
disordered elastic systems. Using them the a.c. transport and
thermodynamic quantities such as the compressibility can be
reliably computed. A pinned quantum crystal exhibit glassy
properties, in a way similar to the Bragg glass for classical
systems (quantum Bragg glass). There is good agreement with the
predicted a.c. transport and the observed behavior of a 2DEG under
a strong magnetic field, making a strong case for a Wigner crystal
phase in such systems. The compressibility is found to be negative
both for a pure Wigner crystal and in the presence of disorder,
and detailed comparison with experiments on that point would be
clearly fruitful. One important point is that the elastic
description is an excellent starting point to obtain the a.c.
properties of such systems. Indeed, although in $d=2$ topological
defects in the crystal (dislocations etc.) are generated by
disorder, the distance at which unpaired defect appear is very
large compared to the scale at which the effects of disorder
manifests themselves. This makes the a.c. transport a probe of
choice to decide on the presence of a Wigner glass phase, since
direct imaging techniques are not yet at the point where the
crystalline structure can be seen.

The d.c. transport is a much more difficult question because of
the need to explicitly take the presence of topological defects
into account. There are however some answers that can be obtained
when the crystal is set in motion at finite velocity. In that case
the defects become less important, and the properties are again
crucially dependent on the periodicity of the system (Moving glass
phase). As was shown in
\cite{giamarchi_moving_prl,ledoussal_mglass_long}, periodic moving
periodic systems have a quite specific dynamics. Indeed due to the
existence of the periodicity in the direction {\it transverse} to
the direction of motion, the motion cannot average completely over
the disorder. The moving system is thus submitted to a random
potential, which leads to a channel like motion. The very
existence of such channels has an important consequence if one
tries to make the system move in the {\it transverse} direction.
Indeed although the particles do move along the channels, the
channels themselves are pinned. This means that even above the
longitudinal threshold field, if one tries to apply a force in the
transverse direction a transverse critical force still exists. For
the two dimensional electronic system, putting a magnetic field is
a simple way of applying a transverse force. If the lattice is
sliding at velocity $v$, it is submitted to a Lorentz transverse
force $F_L = e v B$. The existence of the transverse critical
force $F_{\rm tr}$ thus implies that the channel structure should
not slide as long as $F_L < F_{\rm tr}$. There will thus be no
hall voltage generated. On the other hand when $F_L > F_{\rm tr}$
the channel structure should slide and a Hall voltage exists. The
{\it periodicity} of the crystalline structure thus implies that
one needs a {\it finite} longitudinal current before a Hall
voltage exists. Experiments
\cite{perruchot_wigner_transverse_force} have shown that a  finite
longitudinal current is clearly needed to develop a Hall voltage,
in good agreement with the existence of a transverse threshold.
The existence of such an effect is another direct probe of the
crystalline (existence of a transverse periodicity) nature of the
phase.

We see that there are thus efficient, if not easy, ways via
transport to check for the presence of a pinned crystalline phase
(electronic glass). Most of the experiments discussed in the
preceding sections were for a 2DEG under a strong magnetic field.
It would of course be extremely interesting to use the same
techniques to probe the 2DEG in the absence of the magnetic field,
and analyze the experiments in the line of what was discussed
above to check for the existence of a Wigner crystal in these
systems. Among the interesting possible experiments one can note:
\begin{itemize}
\item Measurement of the optical conductivity. In particular the
density dependence of the pinning peak can be directly checked
against the theoretical predictions of the pinned Wigner crystal.

\item I did not dwell too much of the non-linear transport and on
the existence of a threshold field (as shown in
\fref{fig:wignerfirst}), but a threshold field is also a
characteristics of a pinned crystal. Since it is controlled by the
Larkin length (\ref{eq:lor}) it has to be related to the pinning
frequency. So a comparison between the threshold field in the d.c.
transport and the pinning frequency is another probe of the pinned
Wigner crystal.

\item The Hall tension vs the longitudinal current (i.e. the
measure of the transverse pinning force), as I briefly discussed
here.

\item Although not discussed in these notes, noise measurements
are also a good way to probe the periodic nature of the systems
(see e.g. \cite{togawa_mglass_noise} and references therein).
\end{itemize}

The electronic glasses are thus fascinating objects to play with,
both theoretically and experimentally. Although we have some
answers there are clearly many more unsolved questions than
answers at the moment. Here are some possible directions to
explore in the future in this field
\begin{itemize}

\item What is the d.c. transport? To solve this question one needs
first to understand the behavior of topological defects in a
quantum crystal. Even for classical crystals this question is
open. At the moment there is only an answer for Luttinger liquids
($d=1$) where topological defects do not play a role
\cite{giamarchi_quantum_pinning,nattermann_temperature_luttinger}.

\item In connection with the d.c. transport, what is the out of
equilibrium dynamics of such systems. What is the $I-V$
characteristics. This is quite general question for pinned
structures. For lack of space I will not discuss the question of
the moving crystals here but one can find more details and further
references in
\cite{giamarchi_vortex_review,nattermann_vortex_review} for
classical crystals and in \cite{giamarchi_wigner_review} for the
Wigner crystal.

\item Since the system is a glass, he should show some of the
properties that have been observed other types of glasses. In
particular glasses show aging \cite{cugliandolo_aging_review},
i.e. the result of a measurement depends on the time at which the
measurement is done (the origin of time is fixed by the time at
which the glassy phase has been formed), since a glass relax very
slowly towards its ground state. The question of quantum aging is
in infancy \cite{cugliandolo_dynamics_quantum_aging} and the
experimental consequences for electronic glasses have hardly been
worked out.
\end{itemize}

\section{Acknowledgements}

The bulk of the work discussed in these notes results from a
fruitful and enjoyable collaboration with R. Chitra and P. Le
Doussal, both of whom I would like to specially thank. I would
also like to thank E. Abrahams, A. Yacoby and C.M. Varma for many
interesting discussions on the compressibility in charged systems
and F.I.B. Williams for many enlightening discussions on the
Wigner crystal. I would also like to thank the ITP, where part of
these notes were completed, for hospitality and support under NSF
Grant No. PHY99-07949.


\end{document}